\documentclass[12pt,preprint]{aastex}
\usepackage{amsmath}
\usepackage{lscape}
\usepackage{comment}
\begin{document}

\title{YSO search toward the boundary of the Central Molecular Zone with near-infrared polarimetry}
\author{Tatsuhito Yoshikawa\altaffilmark{1}, Shogo Nishiyama\altaffilmark{2,3}, 
Motohide Tamura\altaffilmark{4,5}, Jungmi Kwon\altaffilmark{6,7}, \\ and Tetsuya Nagata\altaffilmark{8}}
\altaffiltext{1}{Department of Astronomy, Graduate School of Science, Kyoto University, Kyoto 606-8502, Japan; 
yosikawa@kusastro.kyoto-u.ac.jp}
\altaffiltext{2}{National Astronomical Observatory of Japan, Mitaka, Tokyo 181-8588, Japan;
shogo.nishiyama@nao.ac.jp}
\altaffiltext{3}{Miyagi University of Education, Sendai 980-0845, Japan}
\altaffiltext{4}{Department of Astronomy, University of Tokyo, Mitaka, Tokyo 181-0015, Japan}
\altaffiltext{5}{Extrasolar Planet Project Office, National Astronomical Observatory of Japan, Mitaka, Tokyo, 181-8588, Japan}
\altaffiltext{6}{National Astronomical Observatory of Japan, Mitaka, Tokyo 181-8588, Japan}
\altaffiltext{7}{School of Science, The University of Tokyo, Tokyo 113-0033, Japan}
\altaffiltext{8}{Department of Astronomy, Graduate School of Science, Kyoto University, Kyoto 606-8502, Japan; 
nagata@kusastro.kyoto-u.ac.jp}

%\email{yosikawa@kusastro.kyoto-u.ac.jp}
%\email{nagata@kusastro.kyoto-u.ac.jp}
%\email{shogo.nishiyama@nao.ac.jp}

%%%%%%%%%%%%%%%%%%abstract%%%%%%%%%%%%%%%%%%%%
\begin{abstract} 
We have carried out near-infrared polarimetry toward the boundary of the Central Molecular Zone, 
in the  field of ($-1\fdg4 \lesssim l \lesssim -0\fdg3$ and  $ 1\fdg0 \lesssim l \lesssim  2\fdg9, |b|\lesssim 0\fdg1$),
using the near-infrared polarimetric camera SIRPOL on the 1.4 m Infrared Survey Facility telescope.
We have selected 112 intrinsically polarized sources 
on the basis of the estimate of interstellar polarization on Stokes $Q/I-U/I$ planes.
The selected sources are brighter than $K_S=14.5$\,mag and have polarimetric uncertainty $\delta P<1\,\%$.
Ten of these distinctive polarized sources are fit well with spectral energy distributions of young stellar objects 
when using the photometry in the archive of the Spitzer Space Telescope mid-infrared data.
However, many sources have spectral energy distributions of normal stars suffering heavy interstellar extinction; 
these might be stars behind dark clouds.
Due to the small number of distinctive polarized sources and candidates of young stellar object, 
we cannot judge if there is a decline of them outside the Central Molecular Zone.
Many of massive candidates of young stellar object in the literature have only small intrinsic polarization.
This might suggest that their masses are 4--15 M$_{\sun}$,
whose intrinsic polarization has been expected to be small.
\end{abstract}
%%%%%%%%%%%%%%%%%%%%abstract%%%%%%%%%%%%%%%%%%%%

%%%%%%%%%%%%%%%%%%%%introduction%%%%%%%%%%%%%%%%%%%%
\section{Introduction}
As the nearest galactic nucleus, our Galactic Center (GC) provides us with an opportunity 
to study in detail star formation processes in the extreme conditions at the centers of galaxies.
The inner few hundred pc of the GC, known as the Central Molecular Zone (CMZ),
contains several $10^7$M$_{{\sun}}$ of molecular gas \citep{pie00, fer07, lon12}, 
and hosts several prominent star forming regions (e.g., Sgr A, Sgr B2, and Sgr C).
Recent spectroscopic surveys toward nearby galaxies 
reveal the presence of molecular gas and signs of the star formation in their center \citep{maz13}.
Studies of the CMZ are interesting not only for insight into our own Milky Way galaxy, 
but also because they can potentially provide a template to general galactic centers.  

The CMZ is characterized to be different from the Galactic disk in several ways, 
such as its chemistry \citep[e.g.,][]{oka05}, strong turbulence \citep[e.g.,][]{mor96},
high temperature \citep[e.g.,][]{mor96,imm12} and density of gas \citep[e.g.,][]{nag07}, 
and strong magnetic field \citep[e.g.,][]{cro10}.
These conditions may lead to abnormal star formation like top-heavy initial mass function \citep[IMF; e.g.,][]{mor96},
and phenomena associeated with it like CH$_3$OH and H$_2$O masers \citep[e.g.,][]{cha13}.
To study the star formation activity and its history in the whole CMZ,
\citet{fig04} have constructed dereddened luminosity functions in the $m_{{\rm F205W}}$ filter 
in the range of $b=-0\fdg15$--$0\fdg3$
using Hubble Space Telescope/ Near-Infrared Camera and Multiobject Spectrometer.
Comparing these luminosity functions to stellar evolution models,
they have concluded continuous star formation 
at a rate of $\sim 0.14$\,M$_{\sun}$/yr for the entire CMZ with its radius of $\sim$200\,pc.
This value is consistent with the recent star formation activity 
that produced the three massive young clusters in the central 50\,pc 
(the Quintuplet cluster, the Arches cluster, and the central cluster).

In order to gain insight into the nature of star formation in the CMZ, 
\citet{yus09} have searched for young stellar objects (YSOs).
They have carried out multi-color photometric observations with the Spitzer Space Telescope,
and have found a few hundred YSO candidates 
in the region covering $-10'<b<10'$ and $-1\fdg4<l<1\fdg4$.
The 60 percent of the YSO candidates are classified as Stage I with the age of $\sim 0.1$\,Myr, 
and their spatial distribution is not localized but rather uniform. 
This means that the star formation occurred in the extended region of the CMZ $\sim 0.1$\,Myr ago.
\citet{yus09} have also examined the star formation history of the CMZ 
with other signs of star formation like $4.5\mu$m object and 24\,$\mu$m flux.
They have shown that the star formation history of the CMZ is intermittent, 
and this is partially confirmed by the study of \citet{mat11}.
\citet{mat11} have shown that the star formation rates for the entire CMZ are 
0.075\,M$_{\sun}$/yr at 25\,Myr ago and $<$0.02\,M$_{\sun}$/yr at 50\,Myr ago, 
using the existence of three classical Cepheid variable stars \citep[see Figure 3 in][]{mat11}. 
Further observations can improve our understanding for the process of the star formation in the CMZ.

In this study, we carried out a YSO search 
toward the boundary of the CMZ ($(l,b)\simeq (-1^{\circ},0^{\circ})$ and $(2^{\circ},0^{\circ})$) 
by linear polarimetric observations.
Some of YSOs are intrinsically polarized due to the scattering of the stellar light by dust grains 
in their circumstellar disk and nonspherical envelope.
The usefulness of this polarimetric approach for YSO identifications 
are confirmed by model calculations \citep{whi92} and observations \citep{tam89,tam05,yud00,per09}.
We select intrinsically polarized stars as YSO candidates.
After that, we check their color to confirm whether they are genuine YSOs or not,
because YSOs have infrared excess due to re-emission from dust grains in their circumstellar disk and envelope.
Using these two kinds of information, we can select better candidates for YSOs.

\begin{comment}
With polarimetric observations, we can also know the direction of magnetic field and the nature of interstellar medium 
using the information of interstellar polarization \citep[e.g.,][]{nis10,hat13}.
Interstellar polarization is caused by non-spherical dust grains aligned by a magnetic field and/or radiative torque
\citep[dichroic extinction; see, e.g., reviews by][]{laz03,laz07}.
Therefore, the angle of interstellar polarization coincides with the direction of magnetic field. 
Moreover, the wavelength dependency of interstellar polarization and polarization efficiency ($P_{\lambda_2}/E(\lambda_1 -\lambda_2)$)
can be the probe to know the nature of magnetic field and dust grains.
Polarimetric observations can examine both the existence of YSOs and the environment around them.
\end{comment}

%%%%%%%%%%%%%%%%%%%%introduction%%%%%%%%%%%%%%%%%%%%

%%%%%%%%%%%%%%%%%%%%observations & data reduction%%%%%%%%%%%%%%%%%%%%
\section{Observations and data reduction}

\subsection{Observations with IRSF}
We conducted near-infrared (NIR) polarimetric observations of the CMZ with the SIRPOL camera. 
SIRPOL consists of a single-beam polarimeter \citep[a half-wave plate rotator unit and a fixed wire-grid polarizer;][]{kan06} 
and the NIR imaging camera SIRIUS \citep[Simultaneous Infrared Imager for Unbiased Survey;][]{nag99,nag03}, 
and is attached to the 1.4 m telescope IRSF (Infrared Survey Facility).
The camera is equipped with three 1024\,pixel$\times$1024\,pixel HAWAII arrays.
This enables simultaneous observations in the $J$ (central wavelength $\lambda_J=1.25\,\mu$m),
$H$ ($\lambda_H=1.63\,\mu$m), and $K_S$ ($\lambda_{K_S}=2.14\,\mu$m) bands
by splitting the beam into the three wavelengths with two dichroic mirrors.
The image scale of the arrays is $0\farcs45$\,pixel$^{-1}$, yielding a field of view of $460''\times 460''$.

\begin{landscape}
\begin{figure}
\centering
\includegraphics[width=21cm,clip]{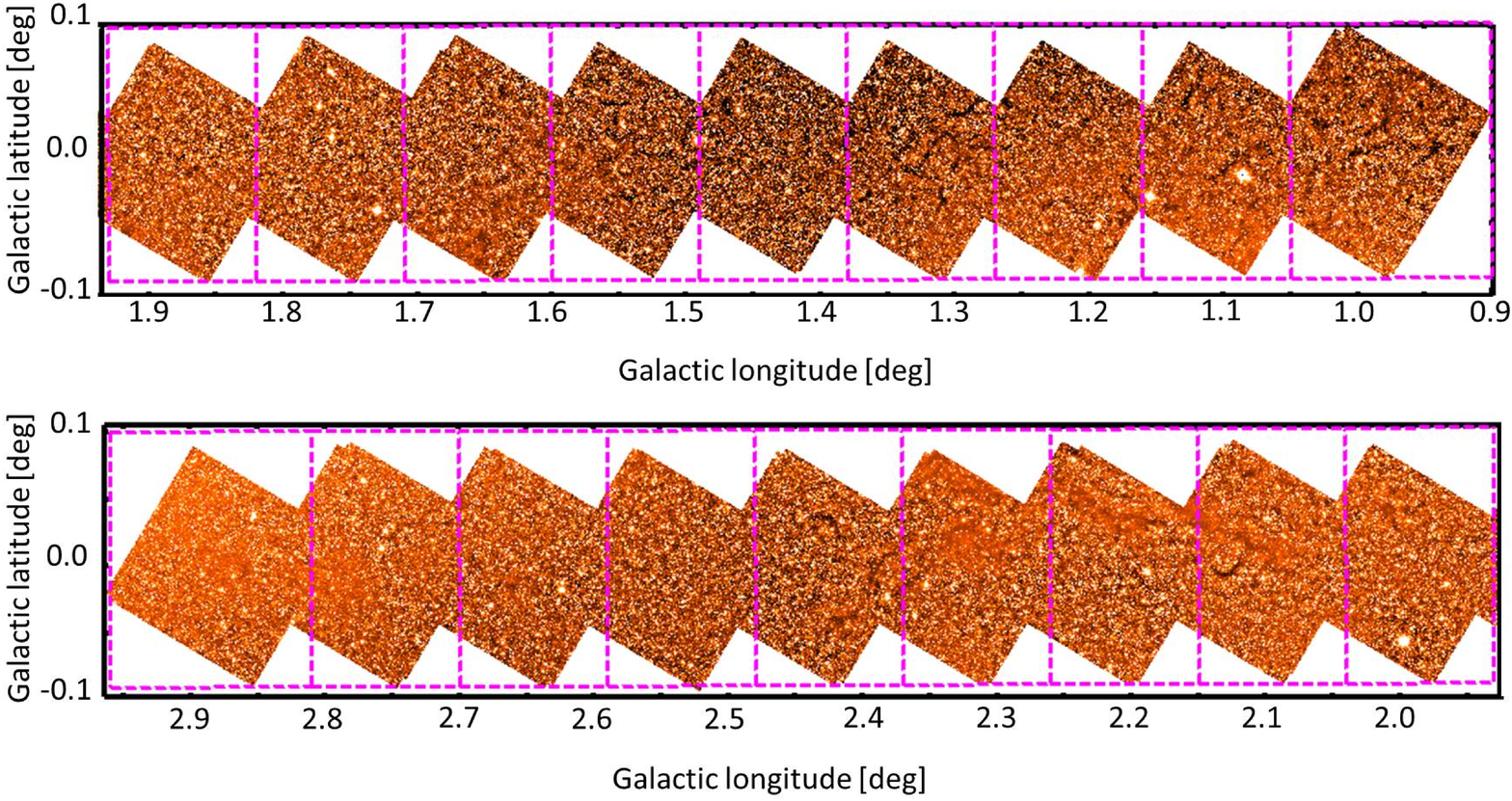}
\caption{Positive longitude side of the $K_S$-band image.
We divide this area into 18 main regions, which are represented by magenta rectangles: 
$l\leq 1\fdg05, 1\fdg05<l\leq 1\fdg16, 1\fdg16<l\leq 1\fdg27, 
\cdots , 2\fdg59<l\leq 2\fdg70, 2\fdg70<l\leq 2\fdg81, 2\fdg81<l$.}
\label{fig:pos_map}
\end{figure}
\end{landscape}

\begin{landscape}
\begin{figure}
\centering
\includegraphics[width=21cm,clip]{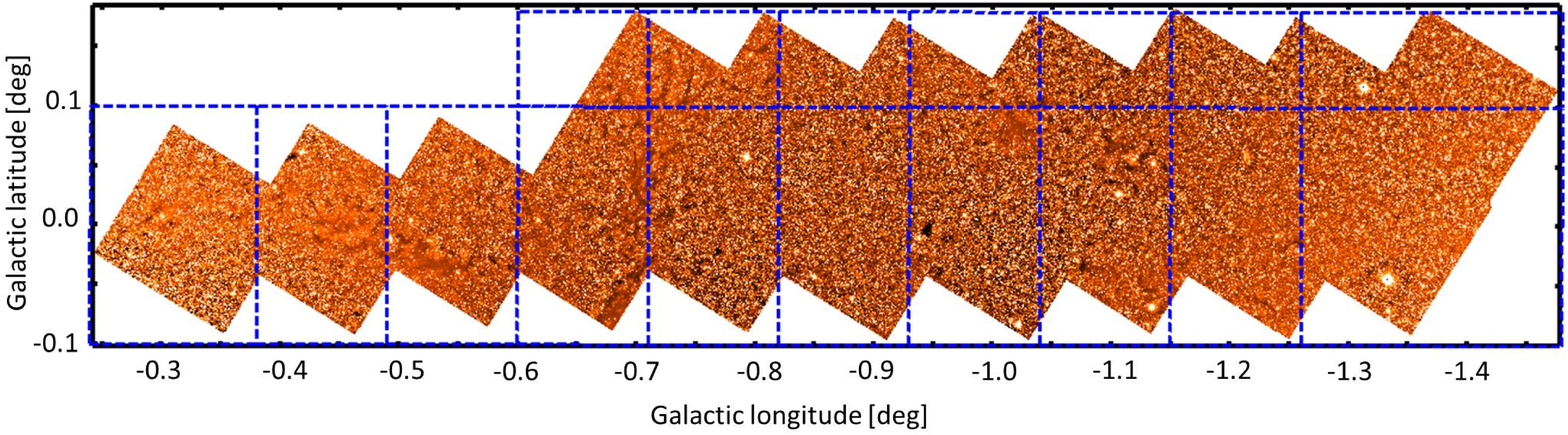}
\caption{Negative longitude side of $K_S$-band image. 
We divide this area into 10 main regions ($-0\fdg1<b\leq 0\fdg1$) and 7 sub regions ($0\fdg1<b$), which are represented by blue rectangles:
$l>-0\fdg38, -0\fdg38\geq l>-0\fdg49, -0\fdg49\geq l>-0\fdg60, 
\cdots , -1\fdg04\geq l>-1\fdg15, -1\fdg15\geq l>-1\fdg26, -1\fdg26\geq l$.}
\label{fig:neg_map}
\end{figure}
\end{landscape}

\begin{landscape}
\begin{figure}
\centering
\includegraphics[width=21cm,clip]{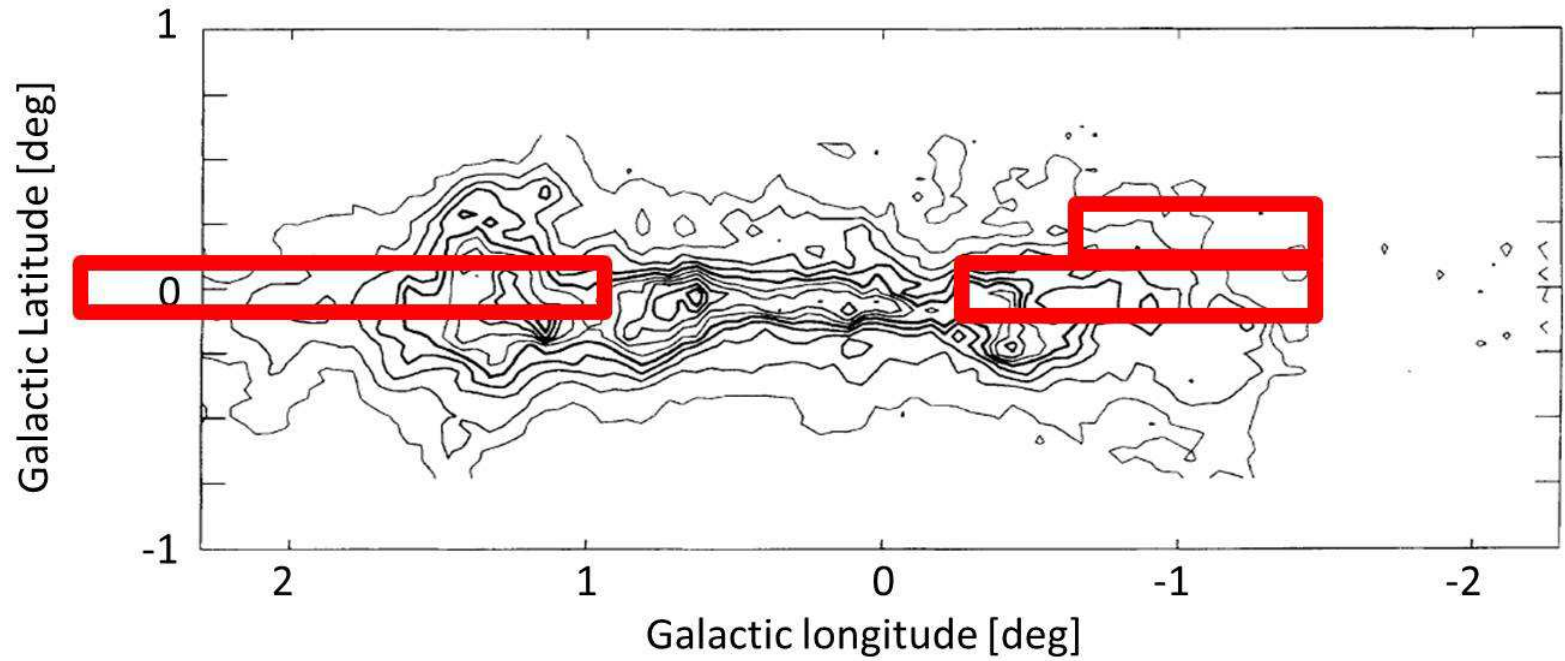}
\caption{Our observational fields shown by red rectangles are superposed on CO (J=1-0) map in \citet{ser96}.
The range of our observational fields are from the high-density region in the CMZ to the outside of the CMZ.}
\label{fig:observational_field}
\end{figure}
\end{landscape}

From 2010 to 2012, we observed 35 fields toward the boundary of the CMZ 
(Figures \ref{fig:pos_map}, \ref{fig:neg_map}, \ref{fig:observational_field}).
The centers of the fields were set at intervals of $400''$ along with the Galactic longitude. %³Šm'É'Í0.11K=396"
We obtained 10 dithered frames on a circle with a radius of $20''$, yielding an effective field of $420''$. 
We performed 10\,s exposures at four wave plate angles ($0^{\circ}, 45^{\circ}, 22\fdg5,$ and $67\fdg5$), 
resulting in a total exposure of 100\,s per wave plate angle for each field. 
We repeated the same observations more than nine times for each field.
Although our observations were carried out on photometric nights, 
seeing depends on sky conditions. 
The range of the seeing is $1\farcs1-1\farcs8$ (FWHM) in the $K_S$ band.

\subsection{Data reduction}
We apply the standard procedures of NIR array image reduction, 
including dark-current subtraction, flat-fielding, self-sky subtraction, and frame combination 
using the IRAF\footnote{Image Reduction and Analysis Facility distributed by the National Optical Astronomy Observatory (NOAO), 
operated by the Association of Universities for Research in Astronomy, Inc. (AURAI) 
and under cooperative agreement with the National Science Foundation (NSF).} software package.
First, we find stars and do photometry with tasks {\it daofind} and {\it phot} in each image. 
For polarimetry, we use stars whose positions are matched within one pixel (= $0\farcs45$) among the images at the four wave plate angles.
Then intensities in the four images are used to calculate the Stokes parameters $I$, $Q$, $U$, 
the degree of polarization $P$, and its position angle $\theta$ using the following equations:
$I = (I_0+I_{45}+I_{22.5}+I_{67.5}) / 2,
Q = I_0-I_{45},
U = I_{22.5}-I_{67.5}, 
P         = \sqrt{(Q/I) ^2+(U/I)^2}$, and $
\theta   = \frac{1}{2}\mathrm{arctan}(U/Q)$, 
where $I_x$ is the intensity with the half wave plate oriented at $x$ deg.  

Photometry of point sources is performed using the DAOPHOT package in IRAF.
We compare aperture photometry with PSF fitting photometry
by calculating $P$ for duplicate sources in overlapping regions of adjacent fields 
observed under different observing conditions.
Aperture photometry gives a better result than PSF fitting photometry which shows systematic offsets.
We adopt the aperture size of $1.0\times$FWHM.
We do the photometry with aperture sizes of 1.0, 1.5, and 2.0 times of stellar FWHM,
and the standard deviations of $P$ are smallest when the aperture size was $1.0\times$FWHM. 
The offsets of $P$ of duplicate sources are consistent 
with the standard deviations of $P$ calculated with $1.0\times$FWHM-aperture photometry.

Since we observed each field more than nine times, 
we calculate the weighted averages and standard deviations of $I, Q,$ and $U$ of each star in all 100\,s-exposure images,
where we use $[1/$(error calculated by task {\it phot} in IRAF)$]^2$ as the weights. 
When we take the average, 
we select the nearest source in other observation images as a matched source, 
and if we do not find any source in a radius of three pixel (= $1\farcs4$), we regard it as non detection.
We remove the bias of $P$ with the equation $P_{{\rm db}} = \sqrt{ P^2 - (\delta P)^2 }$ (\citealt{war74}),
where $P_{{\rm db}}$ is the debiased degree of polarization and $\delta P$ is the photometric error of $P$. 
When we refer to $P$, we do not use sources with $P_{\rm obs}\le \delta P$.

All the data are calibrated for the polarization efficiency of the wave plate and polarizer 
(95.5\,\%, 96.3\,\%, and 98.5\,\% in the $J,H,$ and $K_S$ bands; see \citealt{kan06}).
For photometric calibration, we use the 2MASS Point Source Catalog \citep{cut03} 
in the magnitude ranges of 10.5--12.5\,mag in the $J$ band and 9.5--11.5\,mag in the $H$ and $K_S$ bands, in each field.

For extracting high-quality data, we first check contamination of flux from surrounding sources.
We identify sources within 10\,pixel (= $4\farcs5$) from a source, 
and calculate the sum of contamination flux from the surrounding sources at the peak of the source, 
assuming Gaussian PSF with an FWHM of three\,pixel (= $1\farcs4$).
When the sum of the contamination flux exceeds one percent of the peak flux of the source, 
we remove this source from the list. 
Second, we remove sources around bright and saturated sources to avoid strong contamination of their halo component. 
We obtain positions of bright ($K_S<7$\,mag) sources from the 2MASS Point Source Catalog. 
We checked the size of the halo of the bright sources by eye, 
and determine a circle region in which stars will not be used in the following analysis.
Third, we count how many times IRAF can detect the sources from the (more than nine) 100\,s-exposure images in each field.
When we find stars with {\it daofind}, we set the detection threshold as five times larger than the deviation of local background.
Since our data are obtained under various sky conditions,     
some low-quality sources, which are faint or not separated well from adjacent sources, are not detected by IRAF tasks. 
We regard the sources detected more than seven times in observations in the $K_S$ band as good-quality sources.

\begin{figure}
\centering
\includegraphics[width=15cm,clip]{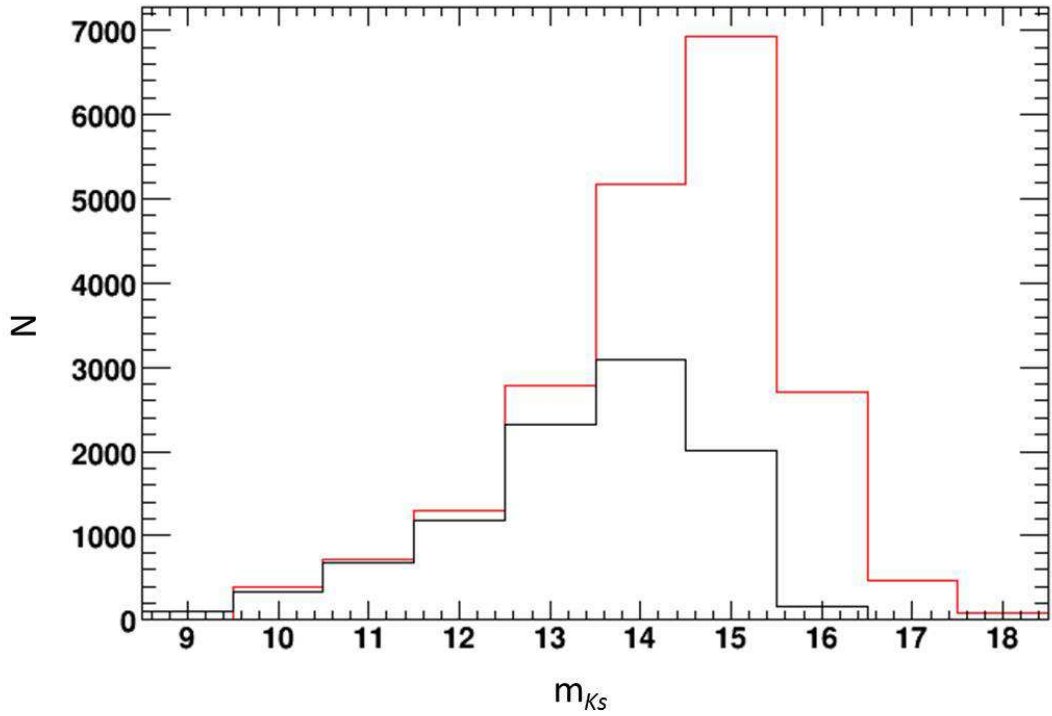}
\caption{$K_S$-band luminosity functions for the good-quality sources (black) and UKIDSS GPS data (red).
We use a region ($-0\fdg9<l<-0\fdg7, -0\fdg1<b<-0^{\circ}$) with no remarkable dark clouds.
The completeness is $\sim 90\,\%$ and $\sim 75\,\%$ in the range of $K_S<13.5$\,mag and $K_S<14.5$\,mag, respectively.}
\label{fig:completeness}
\end{figure}

\begin{figure}
\centering
\includegraphics[width=12cm,clip]{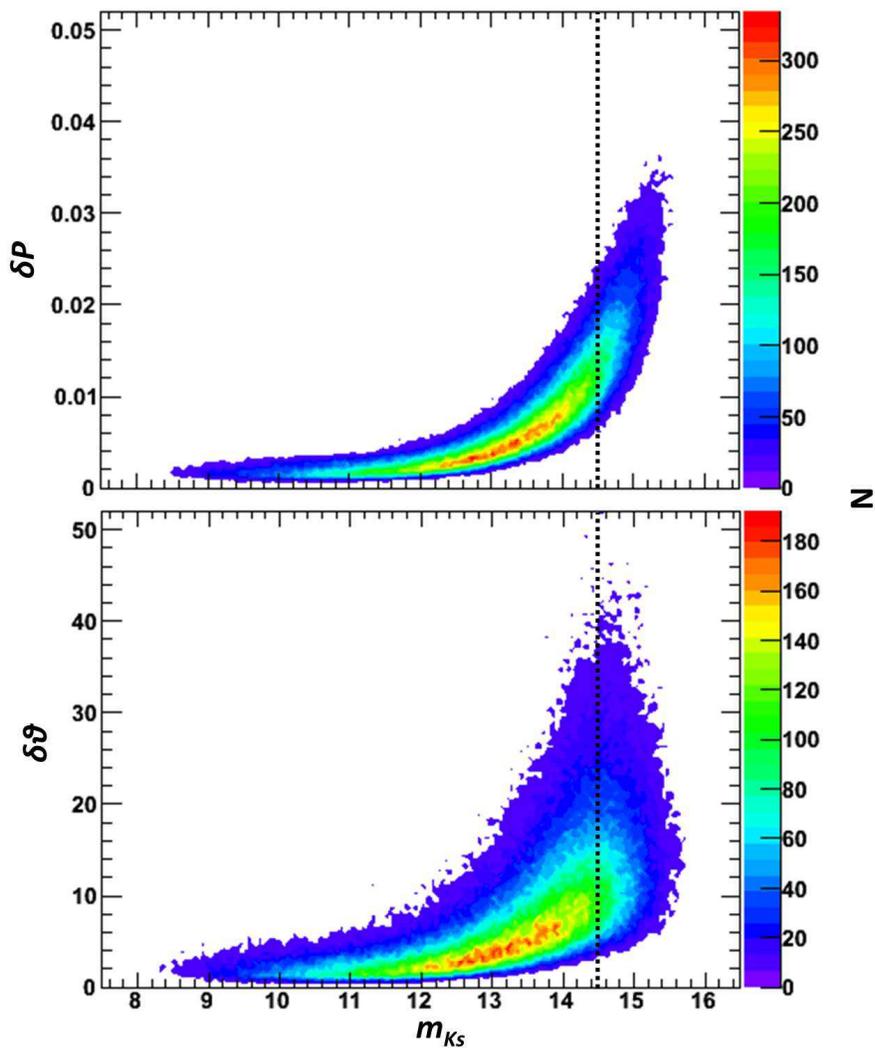}
\caption{Polarization error vs. $K_S$ magnitude for the good-quality sources.
The upper panel exhibits the error of degree of polarization ($\delta P$) vs. $K_S$ magnitude.
The average of $\delta P$ in the range of $K_S=13.5-14.5$\,mag is $<1$\,\% .
The lower panel exhibits the error of polarization angle ($\delta \theta$) vs. $K_S$ magnitude.}
\label{fig:mag_error}
\end{figure}

The completeness of good-quality sources are examined using the UKIDSS GPS point sources
\footnote{UKIDSS and GPS means the United Kingdom Infrared Deep Sky Survey and Galactic Plane Survey, respectively.
The typical spatial resolution of UKIDSS GPS is 0$\farcs$8 (FWHM) in the $K$ band.
For uncrowded regions, the typical completeness at $K\sim18$\,mag is 90\,\% , with uncertainties of $\sim0.2$\,mag.
For crowded regions, the survey is much less sensitive, and the 90\% completeness is located at $K\sim16$\,mag 
\citep[see Figure 1 of][]{luc08}. } \citep{luc08} 
on the assumption that UKIDSS GPS completeness is nearly 100\,\% at $<15$\,mag. 
As shown in Figure \ref{fig:completeness}, the completeness is $\sim 90\,\%$ and $\sim 75\,\%$ 
in the range of $K_S<13.5$\,mag and $K_S<14.5$\,mag, respectively.
Also, the average of $\delta P$ in the range of $K_S=13.5$--$14.5$\,mag is $<1$\,\% (Figure \ref{fig:mag_error}).
We thus exclude sources fainter than 14.5\,mag, 
and finally use 137123 sources in our analysis below.

%%%%%%%%%%%%%%%%%%%%observations & data reduction%%%%%%%%%%%%%%%%%%%%

%%%%%%%%%%%%%%%%%%%%YSO search%%%%%%%%%%%%%%%%%%%%

\section{Search for YSOs with polarization and color}

\subsection{Removal of foreground stars}
First of all, we have to remove foreground sources.
A color-magnitude diagram is useful to achieve this purpose.
In our observations, due to the strong interstellar extinction toward the CMZ, 
only $\sim 50$\,\% and $\sim 10$\,\% of sources detected in the $K_S$ band 
are detected in the $H$ and the $J$ bands, respectively.
Therefore we also use the data of $H$- and $J$-band magnitudes from UKIDSS GPS data \citep{luc08}, 
and merge them with a matching radius of $1''$.
Since we are able to complement most of the data of the $H$-band magnitude this way, but very few of those of the $J$-band magnitude,
we remove foreground sources using $H$ and $K_S$ color-magnitude diagrams.

\begin{figure}
\centering
\includegraphics[width=15cm,clip]{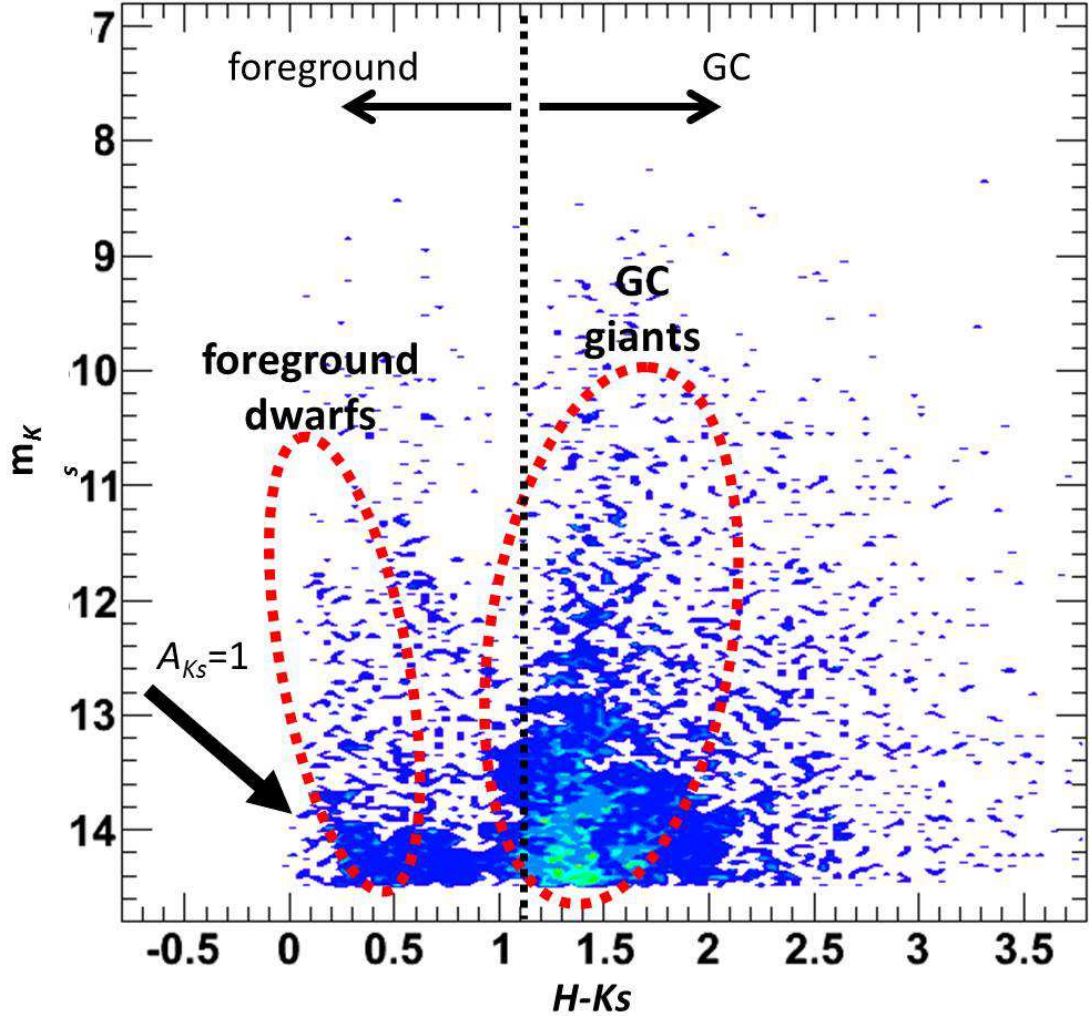}
\caption{Color-magnitude diagram of one main region centered at ($1\fdg22, 0\fdg0$).
The positions of foreground dwarfs and GC giants are roughly represented by red ellipses.
The arrow at the lower left is the reddening vector toward the GC \citep{nis06}. 
For this region, we treat the sources with $H-K_S>1.1$ as the CMZ sources.}
\label{fig:cmd}
\end{figure}

To draw the color-magnitude diagrams, we divide our observational fields into 28 main regions and 7 sub regions 
(see the caption of Figures \ref{fig:pos_map}, \ref{fig:neg_map}).
Since the amount of interstellar extinction toward the CMZ varies in space,
this procedure is necessary. 
The Galactic latitude of the main fields and the sub fields are $-0\fdg1<b<0\fdg1$ and $b>0\fdg1$. 
The Galactic longitude of the centers of both fields are separated in $0\fdg11$.
Note that the sizes of the main fields and the sub fields are different.
We then draw the color-magnitude diagram of each region.
The color-magnitude diagrams show two separated components (Figure \ref{fig:cmd}).
Since there is less interstellar medium between foreground sources and us, 
they show bluer color than the sources located in the CMZ.
The arrow at the lower left in Figure \ref{fig:cmd} is the reddening vector 
in the $H$ and $K_S$ color-magnitude diagram \citep{nis06}.
Dwarfs and giants in the CMZ are strongly reddened along this vector, 
and they are thus located in the redder part.
We define the boundary of the two components by eye for each region, 
and remove the foreground sources.
22770 sources are classified as foreground sources with the color-magnitude diagrams in our observational fields, 
and we treat the remaining 114353 sources as sources located in the CMZ (hereafter the CMZ sources).
Note that sources detected only in the $K_S$ band are included in the CMZ sources,
because they are not detected in the $H$ band probably due to strong extinction toward the sources.

\subsection{Selection of distinctive polarized sources}
\label{sec:polari}
We use a $Q/I-U/I$ plane to select YSO candidates, 
because some of YSOs are intrinsically polarized and show distinctive polarization.
Their positions in the $Q/I-U/I$ plane are apart from those of other sources 
which are intrinsically unpolarized but show polarization due to interstellar polarization.
We aim to extract distinctive polarized sources from the CMZ sources for YSO candidates with the $Q/I-U/I$ plane.

\begin{figure}
\centering
\includegraphics[width=15cm,clip]{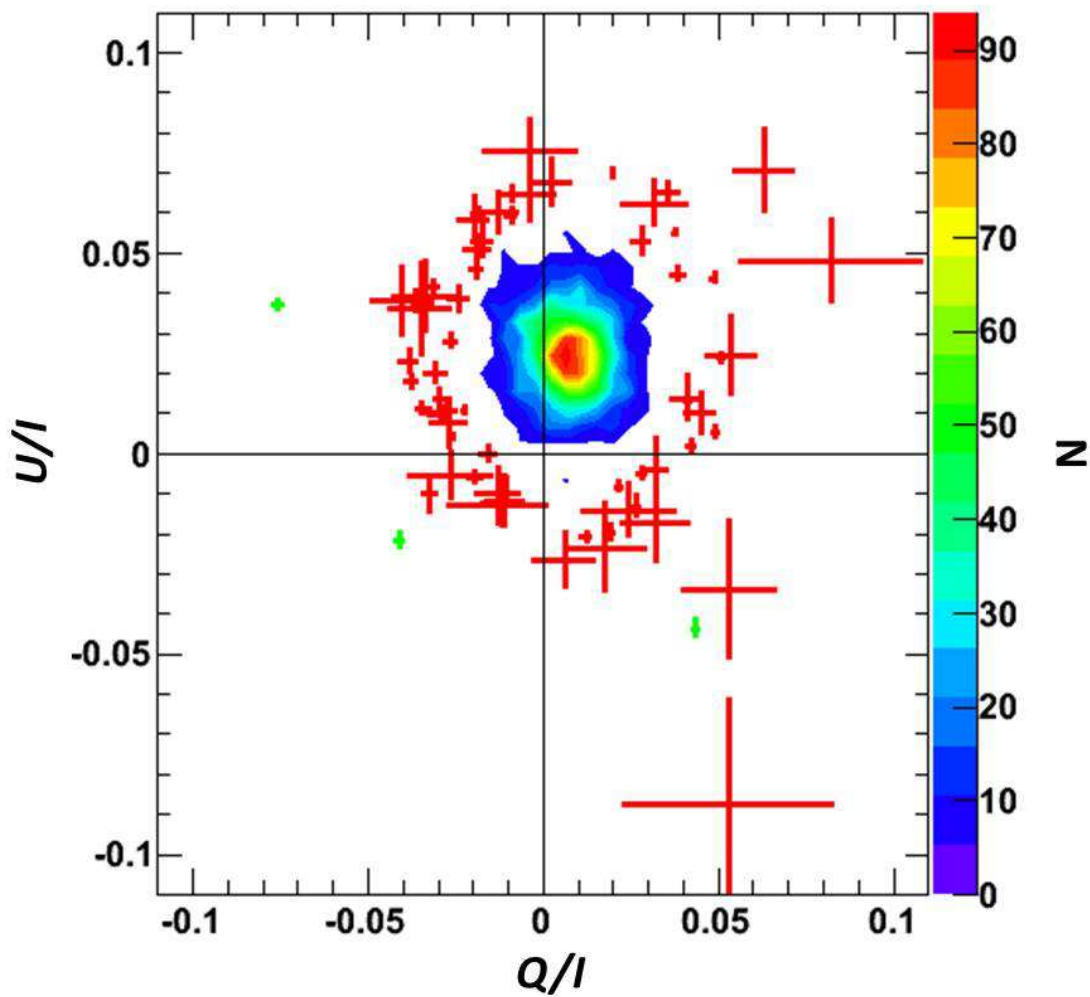}
\caption{$Q/I-U/I$ plane of one main region centered at ($1\fdg22, 0\fdg0$).
The contour is drawn using all good-quality sources in this region (2957 sources) .
Red data points are sources which are far from the peak of contours by 3--5\,$\sigma$,
and green data points are sources which are by $5\,\sigma$, 
where $\sigma$ is the quadratic sum of the polarimetric error of each source 
and the standard deviation determined by Gaussian fitting (Figure \ref{fig:qu_fit}).}
\label{fig:qu_diagram}
\end{figure}

\begin{figure}
\centering
\includegraphics[width=15cm,clip]{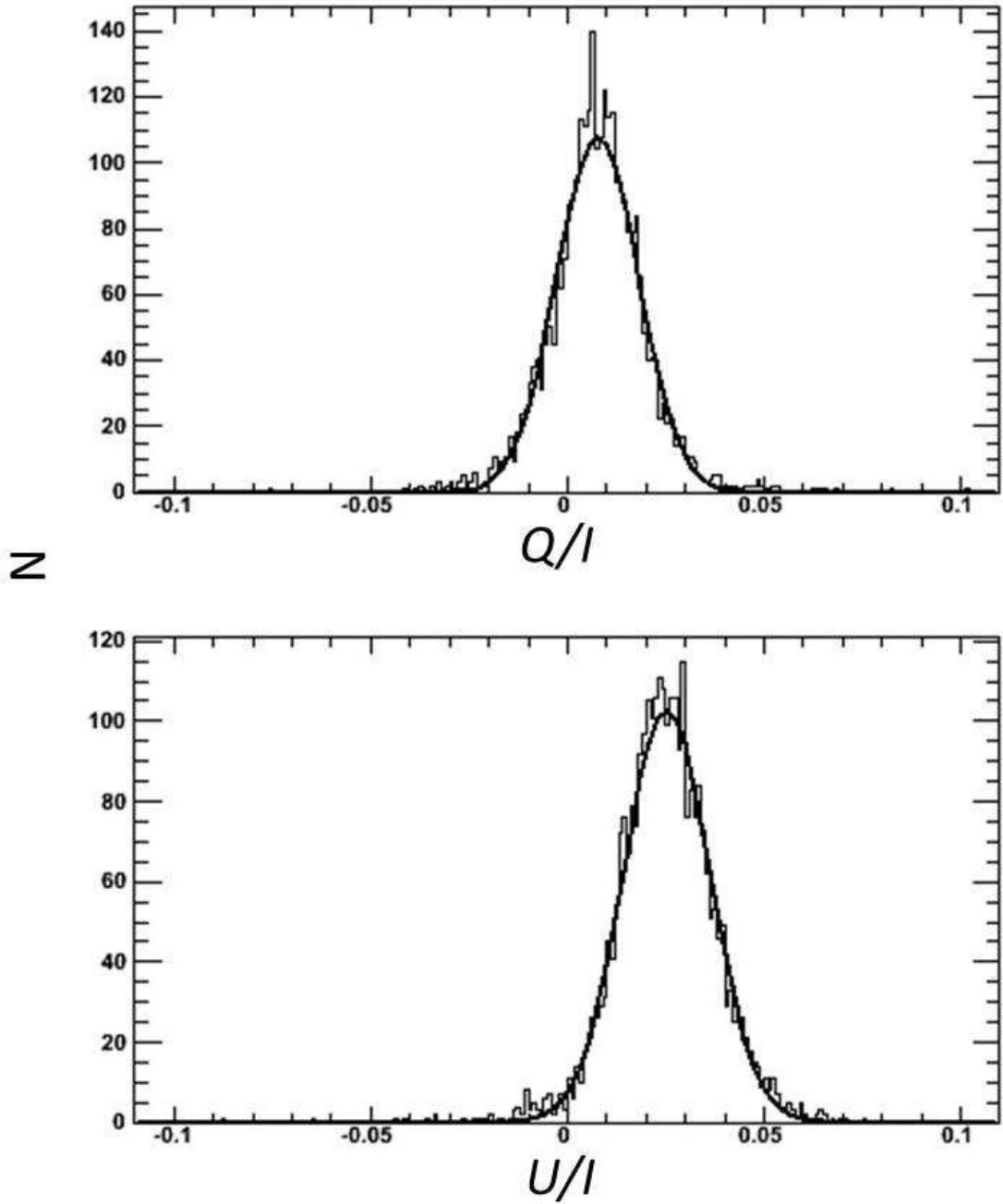}
\caption{Histograms of $Q/I$ (upper) and $U/I$ (lower) for the sources of one main region centered at ($1\fdg22, 0\fdg0$).
Black solid curves are the results of the Gaussian fitting for the histograms.
The standard deviations of these Gaussian functions are used to select distinctive polarized sources in calculating $\sigma$.}
\label{fig:qu_fit}
\end{figure}

First, we draw a $Q/I-U/I$ plane for the CMZ sources in each region (Figure \ref{fig:qu_diagram}).
In Figure \ref{fig:qu_diagram}, the vast majority of sources are concentrated in a well defined region, detached from the origin. 
This detachment reflects interstellar polarization of the region, 
which originates from the dichroic extinction by aligned dust grains along the line of sight.
The spread of the concentrated sources in the $Q/I-U/I$ plane is estimated by fitting with Gaussian functions (Figure \ref{fig:qu_fit}). 
The spread can be attributed to both uncertainties in measurement and real variation of interstellar polarization within the region.
In the $Q/I-U/I$ plane, a number of sources deviate by amounts not explained 
by the errors and the variation in interstellar polarization (green and red plots),
and we classify these sources as candidates of distinctive polarized source.
In selecting them, we calculated the quadratic sum $\sigma$ of the polarimetric error of each source 
and the standard deviation of the Gaussian fitting. 
Green and red plots are apart from the peak of $Q/I$ and $U/I$ by more than $5\,\sigma$ and 3--5\,$\sigma$, respectively.  
In our observational fields, we find 146 ``$5\,\sigma$'' sources (green plots) and 1797 ``3--5\,$\sigma$'' sources (red plots). %1797=1943-146

Second, to improve reliability, we draw another $Q/I-U/I$ plane for the 1000 nearest sources around each $5\,\sigma$ source,
namely, we draw 146 $Q/I-U/I$ planes.
With this procedure, we can estimate more localized interstellar polarization for the 146 sources.
We then find that, 112 sources are apart from the peak of $Q/I$ and $U/I$ by more than $5\,\sigma$ again,
and the other 34 sources are by 3--5\,$\sigma$.
We define these 112 sources as {\it distinctive polarized sources} (see Table \ref{tab:catalogue_I}).
In contrast, we call the other 34 sources and 1797 sources, 
which are selected from the $Q/I-U/I$ plane for each region by 3--5\,$\sigma$, {\it possible distinctive polarized sources.}

\begin{landscape}
\begin{figure}
\centering
\includegraphics[width=21cm,clip]{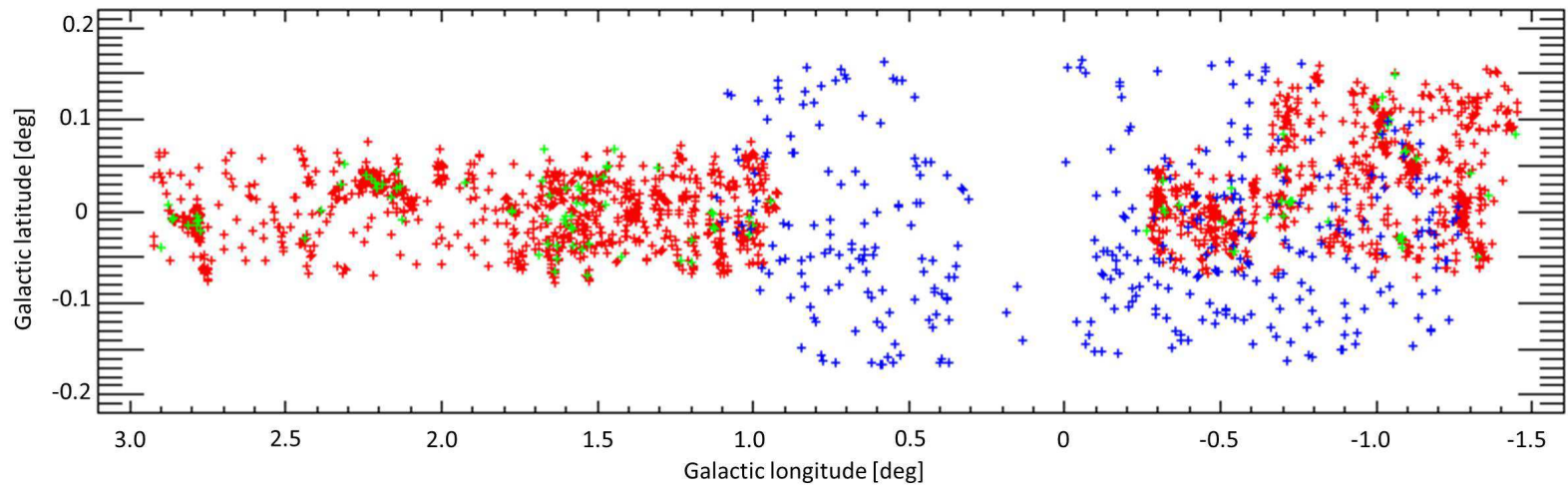}
\caption{Spatial distributions of distinctive polarized sources (green), possible distinctive polarized sources (red), 
and the YSO candidates selected by \citet[blue]{yus09}.
We can see some ``clumps'' of (possible) distinctive polarized sources,
and they reflect the existence of dark clouds.}
\label{fig:spatial_distribution}
\end{figure}
\end{landscape}

\begin{figure}[]
\centering
\includegraphics[width=15cm,clip]{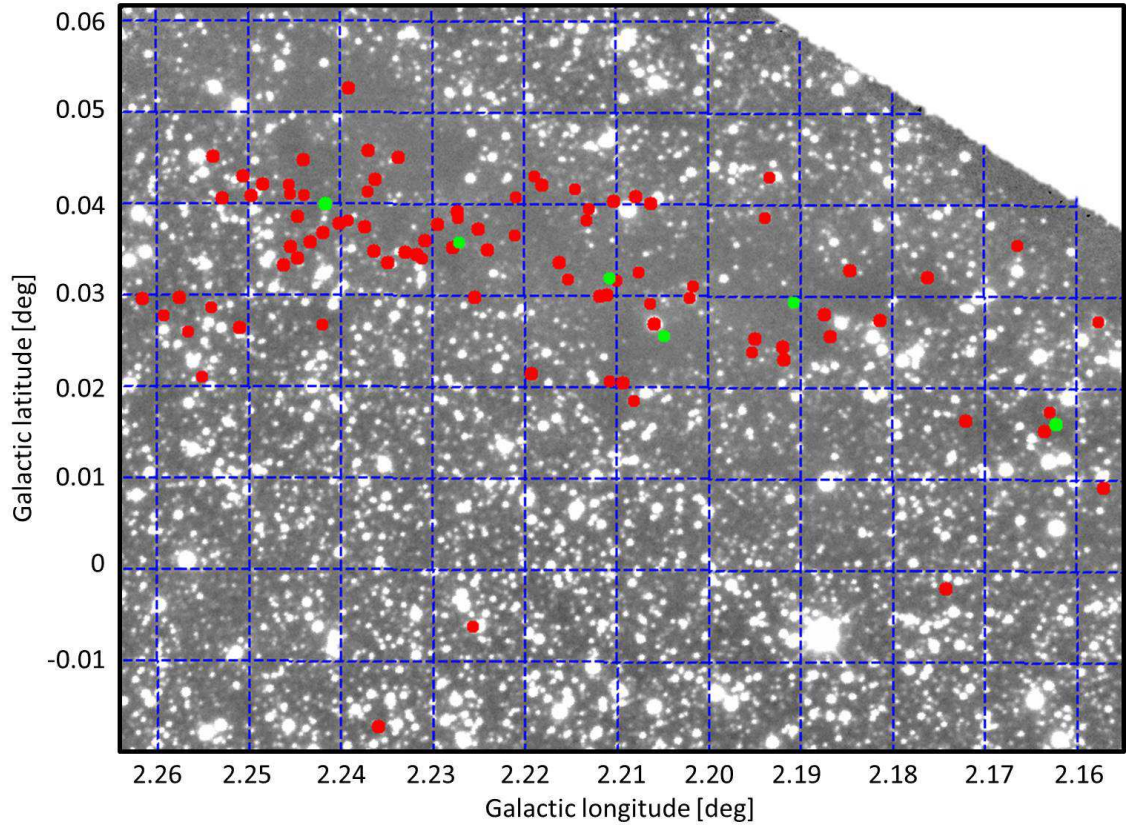}
\caption{$K_S$-band image of a dark cloud.
Red circles and green circles are distinctive possible polarized sources and distinctive polarized sources, respectively.
Many of them are associated with the dark cloud.}
\label{fig:dark_cloud}
\end{figure}

\begin{landscape}
\begin{figure}
\centering
\includegraphics[width=21cm,clip]{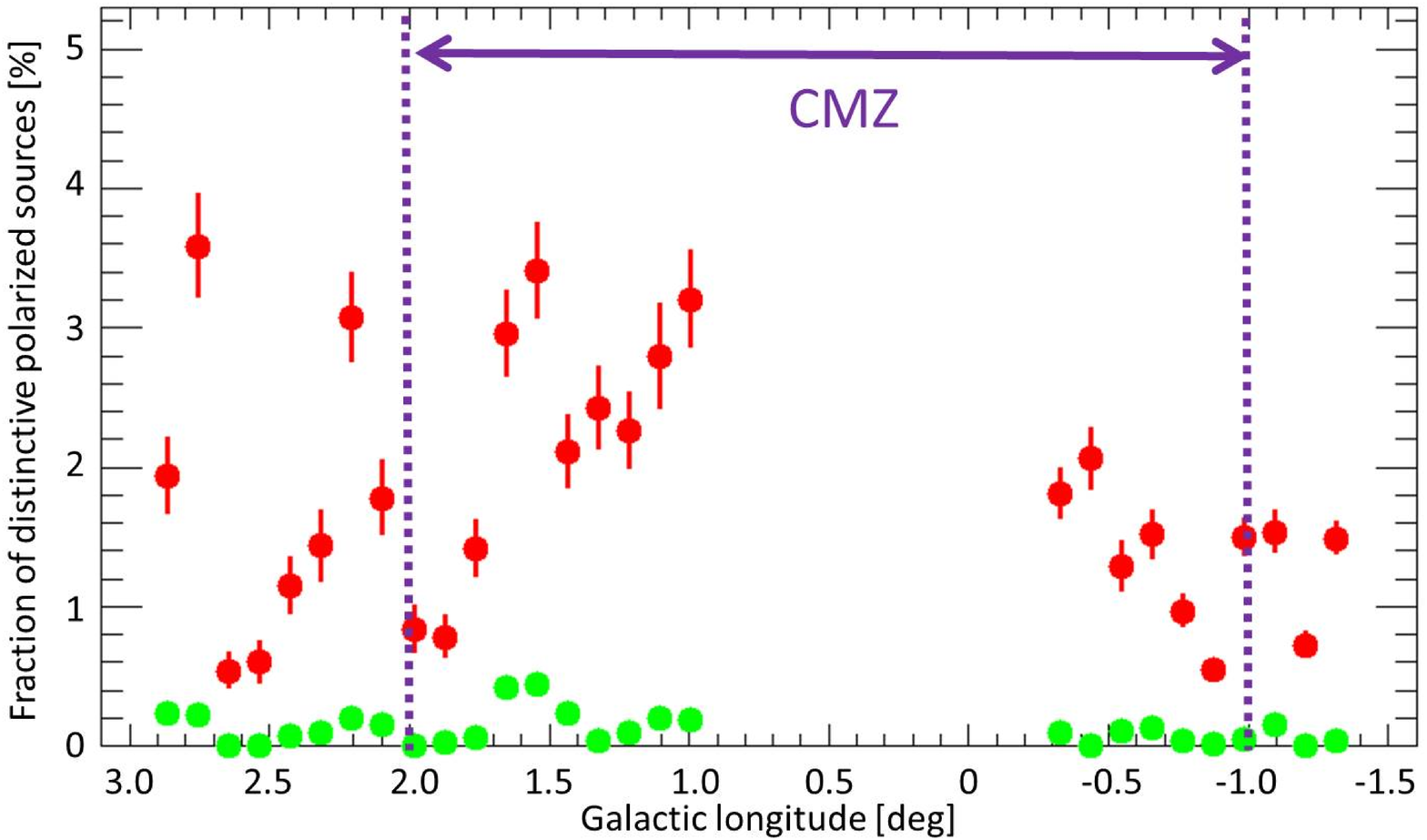}
\caption{Fraction of the polarized sources along the galactic latitude.
Red plots and green plots are the fractions for the possible distinctive polarized sources and distinctive polarized sources, respectively.
Error bars are calculated with poisson error of polarized sources and the CMZ sources.
Purple dot lines represent the boundary of the CMZ \citep[see][]{mor96}.
There seems to be a rapid decline of distinctive possible polarized sources at $\simeq 2^{\circ}$.
The excess of the ratio in the outside of the CMZ comes from some ``clumps'' (see Figure \ref{fig:spatial_distribution}).}
\label{fig:gal_l_distribution}
\end{figure}
\end{landscape}

Figure \ref{fig:spatial_distribution} shows the spatial distributions of the (possible) distinctive polarized sources.
There are some ``clumps'' of (possible) distinctive polarized sources,  
and they are associated with dark clouds in the $K_S$-band images (Figure \ref{fig:dark_cloud}).  
Figure \ref{fig:gal_l_distribution} shows the fraction of the (possible) distinctive polarized sources 
along the Galactic longitude.
The fraction of each region is calculated by dividing the number of the (possible) distinctive polarized sources by that of the CMZ sources.
As evident in the map of molecular gas \citep[e.g.,][]{mor96}, the longitude range of the CMZ is $-1^{\circ}<l<2^{\circ}$.
There seems to be a rapid decline of the {\it possible} distinctive polarized sources at $\simeq 1\fdg8$ and at $\simeq -0\fdg8$ 
in Figure \ref{fig:gal_l_distribution}, although a few points outside the CMZ show larger fractions.
The {\it number} of them also decreased in $l>1\fdg8$.
These longitude points of larger fractions correspond to the ``clumps'' associated with dark clouds.
For the distinctive polarized sources,
we cannot find such a trend due to their small number.

\subsection{Mid-infrared color and SED fitting}
To examine the properties of the distinctive polarized sources, 
especially, to judge whether they are YSOs or not,
mid-infrared observations are effective.
YSOs often exhibit infrared excesses due to thermal emission 
from the circumstellar disks and/or envelopes \citep[e.g.,][]{whi03a,all04}. 
The Spitzer Space Telescope has surveyed the GC with the Infrared Array Camera (IRAC) 
and Multiband Imaging Photometer for Spitzer (MIPS). 
We obtain the data of point sources 
at $3.6\,\mu$m, $4.5\,\mu$m, $5.8\,\mu$m, $8.0\,\mu$m \citep{ram08}, and $24\,\mu$m \citep{hin09} from the archive, 
and we merge the mid-infrared data with our data matching within radii of $1''$ for $3.6\,\mu$m, $4.5\,\mu$m, $5.8\,\mu$m, and $8.0\,\mu$m, 
and $3''$ for $24\,\mu$m.

\begin{figure}
\centering
\includegraphics[width=15cm,clip]{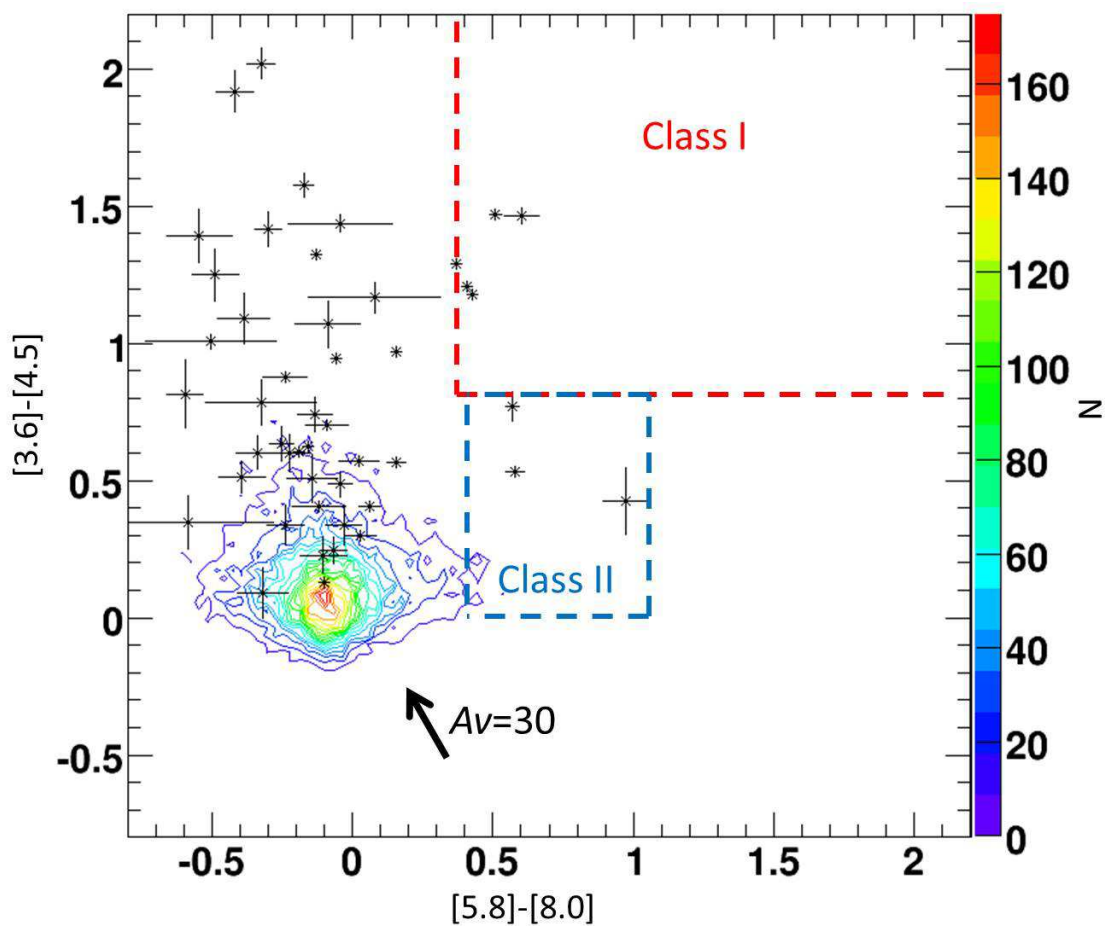}
\caption{Color-color diagram ([3.6]-[4.5] vs. [5.8]-[8.0]) for the distinctive polarized sources.
We plot 49 distinctive polarized sources, which are detected in the four mid-infrared bands.
Contours represent the distribution of colors of the CMZ sources.
Superimposed are the colors of two classes of YSOs calculated by \citet{all04}.
Extinction Vectors, drawn by using the results of \citet{nis08} and \citet{nis09}, are shown for $A_V=30$\,mag.
Five of the distinctive polarized sources are classified as class I YSOs and three of them are class II YSOs.
Remaining sources are located in the region of reddened class III YSOs/normal stars.}
\label{fig:ccd_5sigma}
\end{figure}

We draw a color-color diagram ([3.6]-[4.5] vs. [5.8]-[8.0]) for the distinctive polarized sources in Figure \ref{fig:ccd_5sigma}, 
and compare with the YSO models in \cite{all04}.
Out of the 112 distinctive polarized sources, we can plot 49 sources in the color-color diagram.
Table \ref{tab:matching} exhibits the results of matching between our data and Spitzer data.
Five of them are classified as class I YSOs and three of them are class II YSOs.
The remaining sources are located in the region of reddened class III YSOs/normal stars.

\begin{figure}
\centering
\includegraphics[width=15cm,clip]{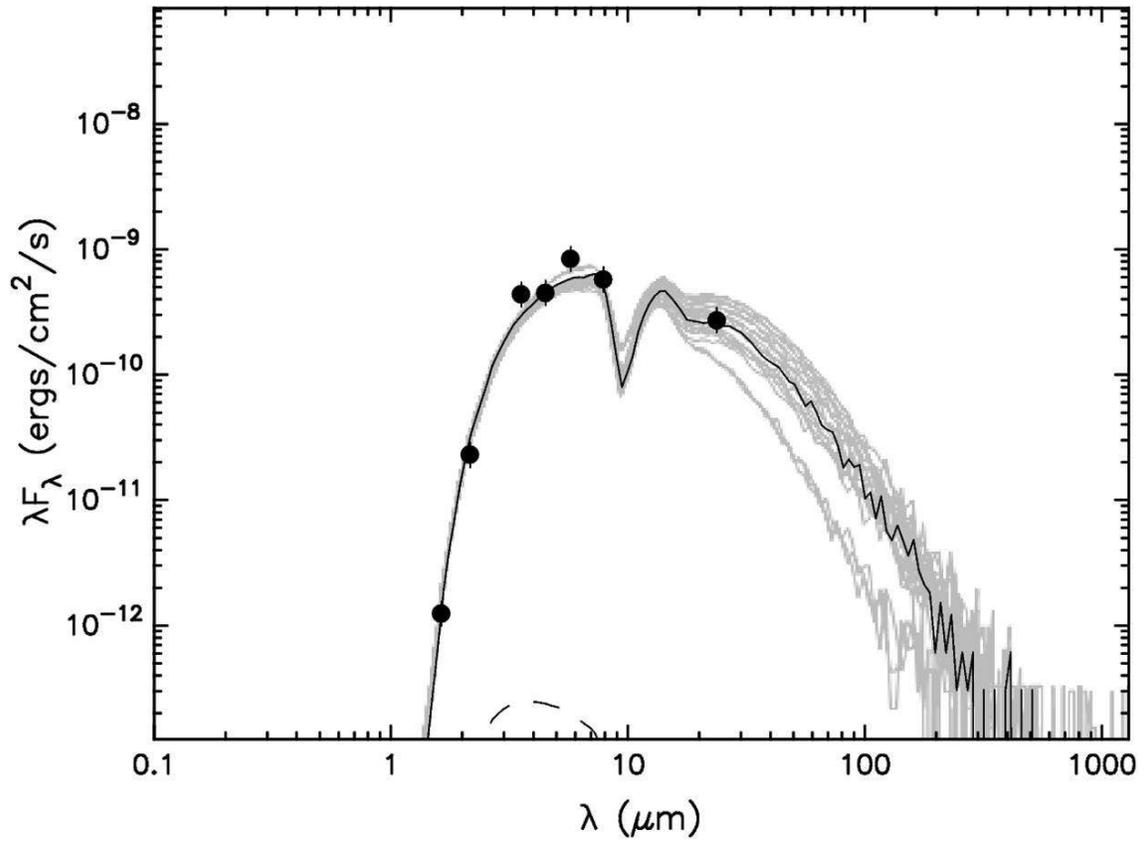}
\caption{SED fit for a distinctive polarized source (\#98).
Black solid line represents the best fit ($\chi^2=7.59$), and gray lines represent good fits with $\chi^2<15$.
For this source, envelope and disk components are necessary.}
\label{fig:SED}
\end{figure}

We also draw SEDs of the distinctive polarized sources.
The SEDs of the sources are analyzed by comparing with a set of SEDs 
produced by a large grid of YSO models \citep{rob06,whi03a,whi03b}.
We use a linear regression fitter \citep{rob07} to find all SEDs from the grid of models 
that are fit within a specified $\chi ^2$ range to the data (Figure \ref{fig:SED}).
In the fitting processes, we set 20\,\% -magnitude errors for all bands 
to account for variability between the IRSF, UKIDSS GPS, IRAC, and MIPS observation dates.
In this SED fitting procedure, at least three-band data are necessary. 
Under the assumption of the distance to the sources of 8.5\,kpc and the interstellar extinction of $A_V=15$--50\,mag, 
75 out of the 112 distinctive polarized sources can be fit well with the models.
Out of the 75 sources, eight sources need disk and/or envelope components to fit their SEDs with the models.
Six of the eight sources are also classified as class I or II YSOs in the color-color diagram (Figure \ref{fig:ccd_5sigma}).
We cannot plot the remaining two sources in the color-color diagram.
The other 67 of 75 sources, for which we carry out SED fitting, can be fit with an only reddened photosphere component.

From the color-color diagram and SED fitting, 
we find 10 YSO candidates, which show both distinctive polarization in the $Q/I-U/I$ plane 
and infrared excess in the color-color diagram and/or SED fitting.
We show their properties in Table \ref{tab:YSO_candidates} and Appendix.
However, most of the distinctive polarized sources can be explained as a normal star with strong interstellar extinction.
We discuss in detail the distinctive polarization which does not seem to come from circumstellar disk/envelope in terms of color information.
 
%%%%%%%%%%%%%%%%%%%%YSO search%%%%%%%%%%%%%%%%%%%

%%%%%%%%%%%%%%%%%%%%nature of distinctive polarization%%%%%%%%%%%%%%%%%%%%

\section{The characteristics of distinctive polarized sources}
Considering color information, we have found 10 YSO candidates from the distinctive polarized sources.
However, the SEDs and colors of many other distinctive polarized sources are better explained without disk/envelope components,
and the others have no color information. 
To know better the characteristics of the distinctive polarized sources,
we examine their intrinsic polarization and how their distinctive polarization is produced.

\subsection{Estimates of intrinsic polarization}
\label{sec:int_pol}
In Figure \ref{fig:ccd_5sigma}, several sources show more than 100\,mag extinction in the $V$ band.
This value is much higher than the average of the extinction toward the GC ($A_V=30$--50\,mag), 
and their distinctive polarization can come from localized interstellar polarization in dark clouds (see Figure \ref{fig:dark_cloud}).
If the distinctive polarized sources suffer strong polarization in general interstellar medium rather than localized polarization, 
they show larger degrees of polarization with similar polarization angles to the typical angle of interstellar polarization. 

\begin{figure}
\centering
\includegraphics[width=15cm,clip]{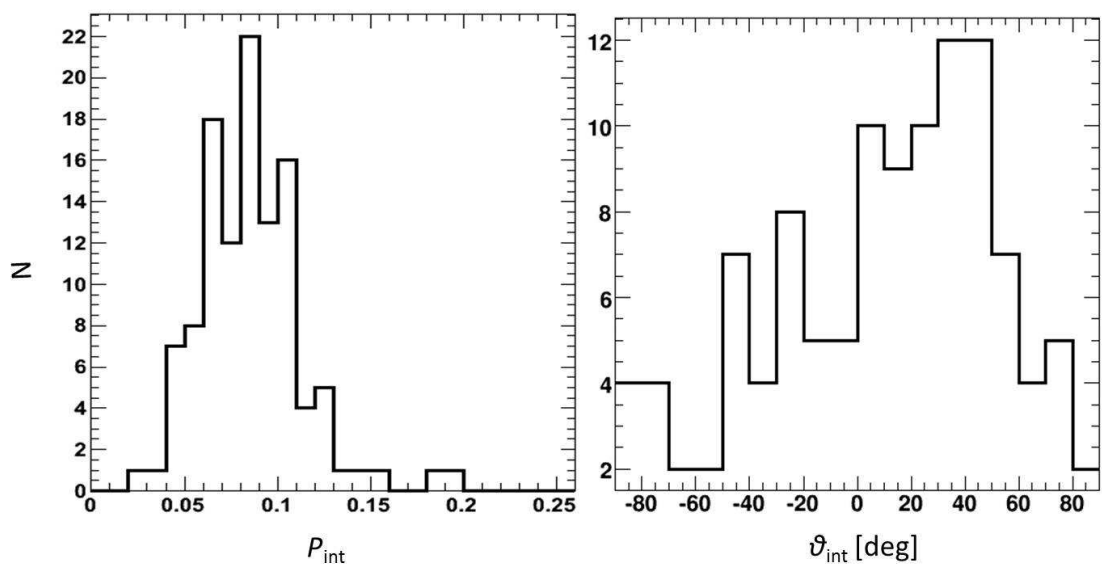}
\caption{Histograms of intrinsic polarization of distinctive polarized sources.
Left panel and right panel exhibit 
the degree of polarization (binning size is 0.01) and polarization angle (binning size is 10$^{\circ}$), respectively.
To calculate the intrinsic polarization, 
we estimate interstellar polarization from 1000 stars around a distinctive polarized source,
and subtract the interstellar polarization from observed polarization of the distinctive polarized source.}
\label{fig:int_pol}
\end{figure}

\begin{figure}
\centering
\includegraphics[width=10cm,clip]{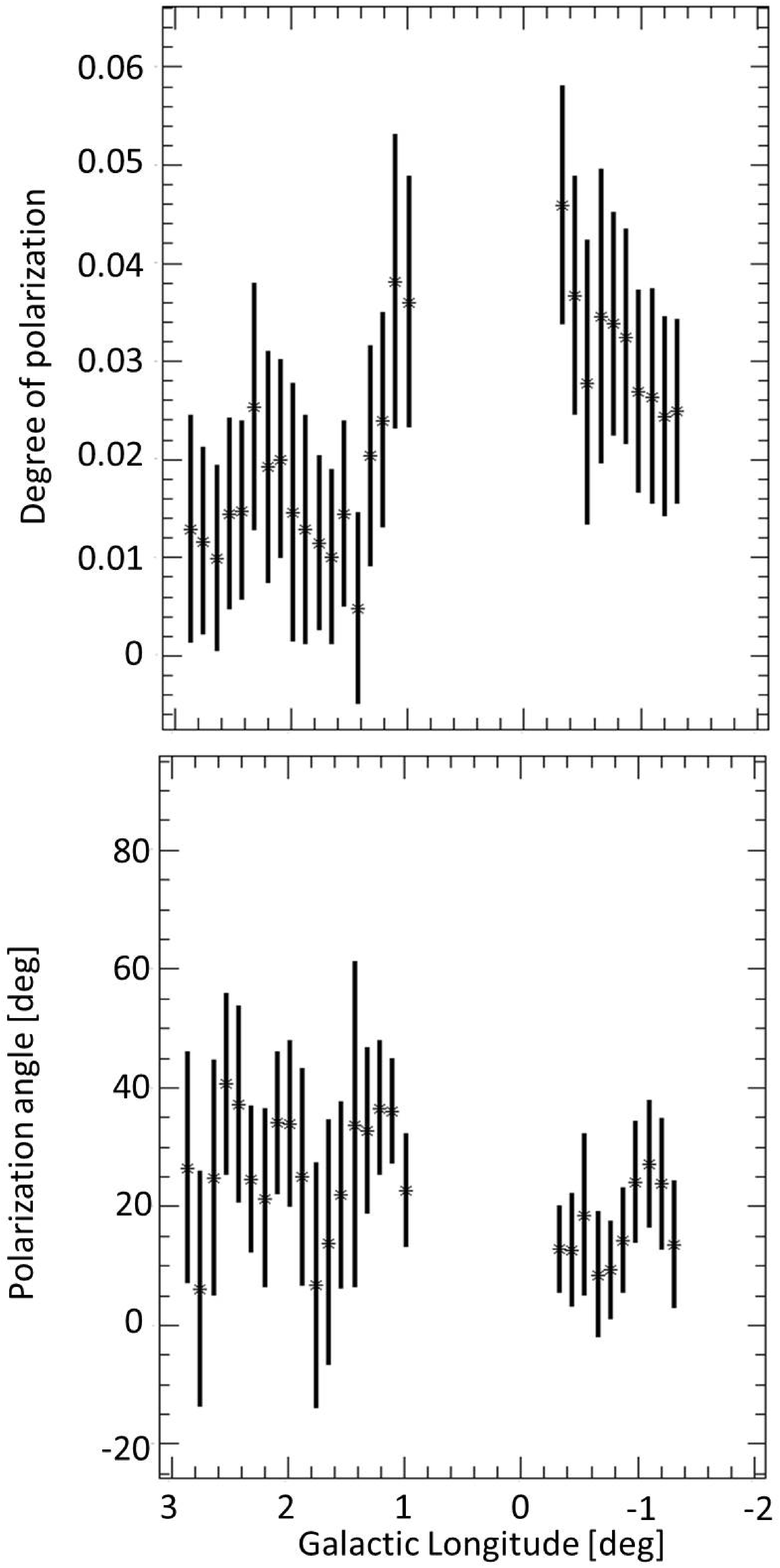}
\caption{Variation of interstellar polarization at $b=0\fdg0$ along the Galactic longitude.
Upper panel is the degree of polarization, 
and lower panel is the polarization angle.
The averages of them are in good agreement with \citet{hat13}.
There seems to be a variation in the degree of polarization.}
\label{fig:ISP_distribution}
\end{figure}

We estimate intrinsic polarization of the distinctive polarized sources 
and draw the histograms in Figure \ref{fig:int_pol}.
To estimate the intrinsic polarization, 
we have calculated the mean interstellar polarization for the nearest 1000 stars around each distinctive polarized source,
and subtract the mean interstellar polarization 
from the observed polarization of each distinctive polarized source as stated in section \ref{sec:polari}.
Some of their intrinsic polarization angles are not consistent 
with a typical polarization angle ($\sim 20^{\circ}$) of interstellar polarization toward the CMZ (Figure \ref{fig:ISP_distribution}).
This means that a part of the distinctive polarization are not explained by the stronger interstellar polarization only.
The different polarization angles would imply that a region of greater extinction like a dark cloud 
where the direction of dust alignment and magnetic fields are different from that of the CMZ, 
contributes to the polarization.

\subsection{Polarization of surrounding stars of each distinctive polarized sources}
\begin{figure}
\centering
\includegraphics[width=15cm,clip]{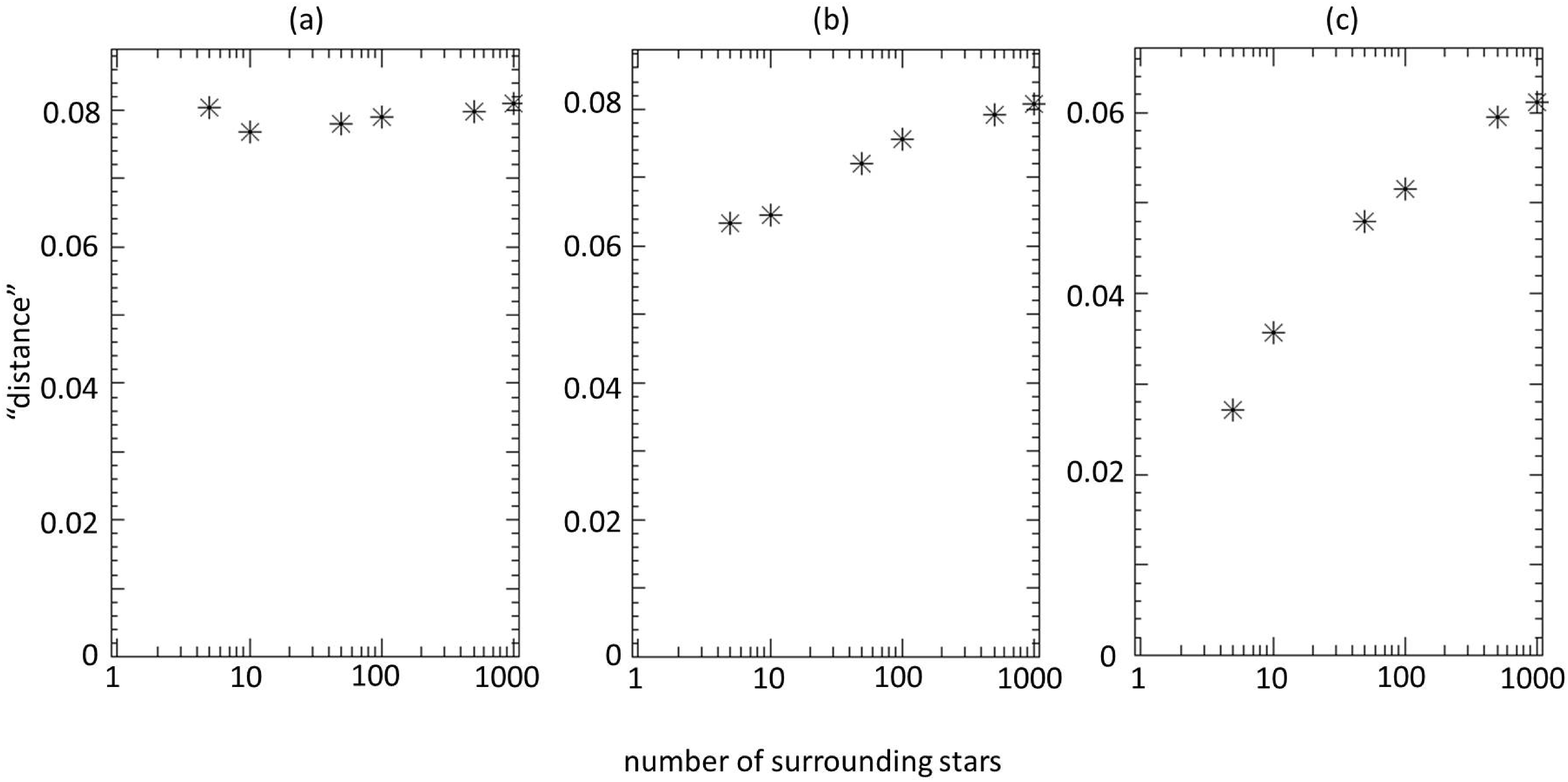}
\caption{Difference of polarization between the distinctive polarized sources and surrounding sources.
The vertical axis represents the ``distance'' in the $Q/I-U/I$ plane 
between one of the distinctive polarized source and the surrounding sources.
The ``distance'' is calculated by $\sqrt{((Q/I)_{{\rm pol}}-<Q/I>_{{\rm sur}})^2+((U/I)_{{\rm pol}}-<U/I>_{{\rm sur}})^2}$, 
where $(Q/I)_{{\rm pol}}$ and $(U/I)_{{\rm pol}}$ are those of the distinctive polarized source,
and $<Q/I>_{{\rm sur}}$ and $<U/I>_{{\rm sur}}$ are the averages of $Q/I$ and $U/I$ of the surrounding sources. 
The horizontal axis represents the number of the nearest surrounding sources $n$, which we use to calculate the average of polarization.
In panel (a) (\#81), the average of polarization of the surrounding sources does not change as the number $n$ decreases.
In panel (c) (\#88), the ''distance'' in the $Q/I-U/I$ plane decreases to 0 as the number $n$ decreases.
Panel (b) (\#86) shows the middle character of these two panels.
For 112 distinctive polarized sources, 63, 19, and 30 sources are classified as the type (a), (b), and (c), respectively.}
\label{fig:pol_diff}
\end{figure}

To examine the nature of the distinctive polarized sources, 
we check how different their polarization are from the adjacent stars.
Figure \ref{fig:pol_diff} is the example of difference of polarization 
between three of distinctive polarized sources and their surroundings. 
The vertical axis represents the ``distance'' 
between a distinctive polarized source and the mean position of surrounding sources in the $Q/I-U/I$ plane.

The ``distance'' is calculated by $\sqrt{((Q/I)_{{\rm pol}}-<Q/I>_{{\rm sur}})^2+((U/I)_{{\rm pol}}-<U/I>_{{\rm sur}})^2}$, 
where $(Q/I)_{{\rm pol}}$ and $(U/I)_{{\rm pol}}$ are those of the distinctive polarized source, 
and $<Q/I>_{{\rm sur}}$ and $<U/I>_{{\rm sur}}$ are the averages of $Q/I$ and $U/I$ of the surrounding sources.
The horizontal axis represents the number of the nearest surrounding sources $n$, 
which we use to calculate the average of polarization.
Drawing these figures for the 112 distinctive polarized sources, 
we find two kinds of behavior in the polarization of surrounding sources.
One shows no or little variation in the polarization 
as the number of the surrounding sources decreases (type (a), shown in the left panel of Figure \ref{fig:pol_diff}).
The other shows great variation (type (c), shown in the right panel of Figure \ref{fig:pol_diff}). 
63 distinctive polarized sources are classified as type (a), and 30 as type (c), 
and the other 19 show intermediate characters of these two types 
(type (b), shown in the middle panel of Figure \ref{fig:pol_diff}).    
The distinctive polarized sources classified as type (a) have very different polarization from their surrounding sources.
Their polarization is different from the average of polarization of even surrounding five stars. 

Of the sources classified as YSO candidates with mid-infrared color or SED fitting,
only one is classified as type (b) and the others are all type (a).
Therefore the YSO candidates clearly show different polarization from their adjacent sources.

In the last section, we have also found that many distinctive polarized sources are not YSO candidates, 
but normal stars strongly reddened by interstellar medium like dark clouds.
For the distinctive polarized sources of type (a), 
dark clouds with a size of $\sim 11''$--$16''$ are necessary to explain the distinctive polarization.
This size is estimated by calculating the distances 
between the distinctive polarized sources and the surrounding five to ten sources. 
Assuming that these dark clouds are in the CMZ with a distance of 8\,kpc, 
the physical size of such dark clouds is 0.44--0.64\,pc.
This size is equivalent to that of a single molecular cloud \citep[e.g.,][]{hey04}. 
We suggest that we may be observing individual molecular clouds in the CMZ through the distinctive polarized sources of type (a) and (b). 
Another possibility is that some distinctive polarized sources are very luminous stars and behind the CMZ,
and thus they suffer from stronger interstellar reddening and polarization.

\subsection{$H-K_S$ color difference}
\begin{figure}
\centering
\includegraphics[width=15cm,clip]{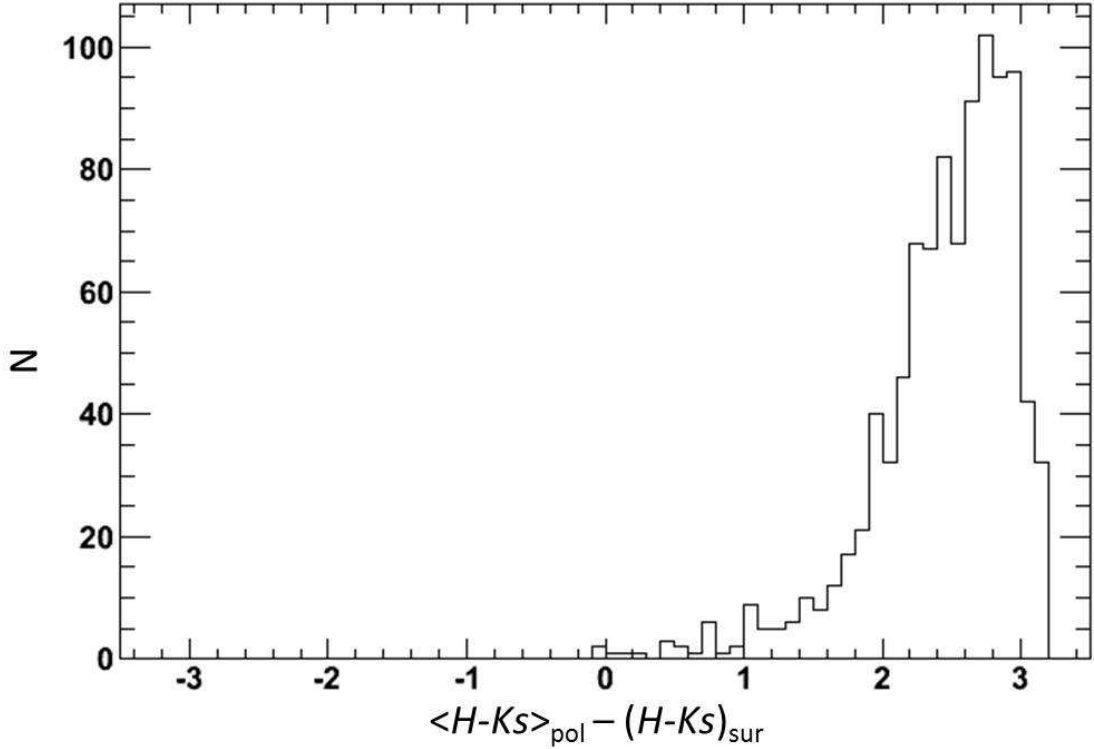}
\caption{Histogram of difference of $H-K_S$ color between one of the distinctive polarized sources (\#1) and surrounding sources.
The horizontal axis represents $(H-K_S)_{{\rm pol}}-(H-K_S)_{{\rm sur}}$, 
where $(H-K_S)_{{\rm pol}}$ is the $H-K_S$ color of the distinctive polarized source
and $(H-K_S)_{{\rm sur}}$ is the $H-K_S$ colors of the surrounding sources. 
In this histogram, a positive value indicates that the distinctive polarized source has redder color than those of the surrounding sources.
Out of the 112 distinctive polarized sources, we obtain the $H-K_S$ color of 81 sources. 
Eleven of them have a similar color (the difference is -0.5 to 0.5) to those of the surrounding sources,  
and the others have redder colors.}
\label{fig:hk_diff}
\end{figure}

We also check if the $H-K_S$ color of the distinctive polarized sources are differenct from the surrounding sources.
Figure \ref{fig:hk_diff} exhibits the difference of observed $H-K_S$ color 
between a distinctive polarized source and its surrounding sources.
Although we use the 1000 nearest surrounding sources, 
some of them are not detected in the $H$ band,
and we remove them from the histogram.
The horizontal axis represents $(H-K_S)_{{\rm pol}}-(H-K_S)_{{\rm sur}}$, 
where $(H-K_S)_{{\rm pol}}$ is the $H-K_S$ color of the distinctive polarized source 
and $(H-K_S)_{{\rm sur}}$ is the $H-K_S$ color of the surrounding sources,   
and a positive value means a redder color of the distinctive polarized sources than the surrounding sources.
Out of the 112 distinctive polarized sources, 
eight sources have similar colors (the difference is -0.5 to 0.5) to that of the surrounding sources,  
and other 80 sources have redder colors.
The other 24 distinctive polarized sources are not detected in the $H$ band.
We calculate the averages and the standard deviations of $(H-K_S)_{{\rm pol}}-(H-K_S)_{{\rm sur}}$ 
using 1000 surrounding sources. 
The results are $1.64\pm 0.84$, $1.98\pm 0.84$, and $1.51\pm 0.87$ 
for the type (a), (b), and (c) in Figure \ref{fig:pol_diff}, respectively.
Thus, the distinctive polarized sources are generally redder than the surrounding sources in any types.
For the sources classified as YSO candidates with mid-infrared color or SED fitting,
one of them has no $H$ band data (\#12), two of them have similar color (\#13 and \#23), 
and the average of the others is $1.88\pm 0.37$.
We do not find any trend in the types (a)-(c) or whether they are classified as YSO candidates.

%%%%%%%%%%%%%%%%%%%%%%%%nature of distinctive polarization%%%%%%%%%%%%%%%%%%%%%%

%%%%%%%%%%%%%%%%%%%%%%%%Yusef YSOs%%%%%%%%%%%%%%%%%%%%%%%%%%%%%%%%%%%%

\section{Polarization of YSO candidates in the literature}
In the CMZ, \citet{yus09} have found a few hundred YSO candidates
\footnote{\cite{an11} have carried out spectroscopic observations for their YSO candidates in the CMZ with SST/IFS
and suggest that the half of the YSOs selected in \cite{yus09} are not YSOs.} 
with multi-color photometry.
They have selected YSO candidates by drawing a color-magnitude diagram ([24] vs. [8]-[24]) 
and by fitting with SED models of \citet{rob06}, 
and have obtained parameters of the candidates. 
In our observational fields, we obtain polarimetric information of 68 YSO candidates of 
\citet[hereafter, we call these sources the Yusef-Zadeh sources]{yus09}.  
They consist of 36 Stage I YSOs, 24 Stage II YSOs, and 8 Stage III YSOs.

To check the polarimetric properties of the Yusef-Zadeh sources, 
we draw $Q/I-U/I$ planes using 1000 stars around the Yusef-Zadeh sources.
Polarizations of 64 Yusef-Zadeh sources are similar to typical values of the surrounding sources within their errors. 
Three Yusef-Zadeh sources are possible distinctive polarized sources 
(SSTGC\footnote{Spitzer Space Telescope Galactic center number based on the IRAC catalogs \citep{ram08}} 
134327 (Stage I), 336047 (Stage II), and 358063 (Stage II)), 
and one source is a distinctive polarized source (SSTGC 861689 (Stage I), \#40 ).

Under certain conditions, YSOs show observable polarization in infrared wavelengths \citep[see Figure 9 in][]{rob06}.
For Stage I YSOs, most of them with a mass of $\sim2$\,M$_{{\sun}}$ 
show polarization around $2\mu$m except for the face-on view of disk/envelope.
For Stage I/II YSOs with a mass of $\sim20$\,M$_{{\sun}}$, 
polarization is detectable only for the disk/envelope seen almost edge on.
According to \citet{yus09}, their sources within our observational fields 
are $\sim4$--15\,M$_{{\sun}}$ (the median mass of $\sim 7$\,M$_{{\sun}}$),
and 36 of them are Stage I YSOs on the basis of the SED fitting.
Considering that only four of them exhibit $>3\,\sigma$ polarization, 
it seems that $\sim 7$\,M$_{{\sun}}$ Stage I YSOs do not exhibit large polarization as $\sim 20$\,M$_{{\sun}}$ YSOs do not.

The small rate of large polarization in Yusef-Zadeh sources is still surprising.
The left panel of Figure \ref{fig:int_pol} exhibits the degree of intrinsic polarization of the distinctive polarized sources,
which indicates that $P_{{\rm int}}>3$\,\% sources are identified as the distinctive polarized sources in our analysis.
\citet{yud00} have shown that out of 13 Herbig Ae/Be stars whose intrinsic polarization he found in the literature, 
three have $P_{{\rm int}}>3$\,\% in the $V$ or $R$ band \citep[see Table 3 in][]{yud00}. %ãn'Ō덷'ð'Æ'é'Æ50%}35%
Moreover, \citet{per09} have shown that for the degree of intrinsic polarization in the $H$ band, 
one of seven Herbig Ae/Be stars exceeds 3\,\%. 
These results suggest that $\sim10$--20\,\% of Herbig Ae/Be stars show $P_{{\rm int}}>3$\,\%.
Although the statistics of NIR polarimetry for YSOs needs to be improved, 
such a solid detection rate is not consistent with the low detection rate of polarization of $>3$\,\% in Yusef-Zadeh sources.
This inconsistency may come from the selection bias of Yusef-Zadeh sources, which are red in the color of [8]-[24].
More accurate observations and analysis can probably detect the intrinsic polarization of YSOs.

%%%%%%%%%%%%%%%%%%%Yusef YSOs%%%%%%%%%%%%%%%%%%%%%%%%

%%%%%%%%%%%%%%%%%%%%summary%%%%%%%%%%%%%%%%%%%%

\section{Summary}
In this paper, we have studied the intrinsic polarization of bright ($K_S<14.5$\,mag) infrared sources toward the outer part of CMZ.
We have selected 112 distinctive polarized sources,
but only ten of them are good YSO candidates when we use color-color diagrams 
and SED fits of mid-infrared photometry with the Spitzer Space Telescope.
Furthermore, the YSO candidates in Yusef-Zadeh et al. (2009) have in general only small intrinsic polarization.
Therefore, near-infrared polarimetry might not be very efficient in identifying YSOs near the CMZ,
but the selected YSO candidates and several other distinctive polarized sources are good targets of further detailed studies.

%%%%%%%%%%%%%%%%%%%%summary%%%%%%%%%%%%%%%%%%%%

\acknowledgments
This work was partly supported by the Grant-in-Aid for JSPS Fellows for young researchers (T.Y.).
This work was also supported by KAKENHI, 
Grant-in-Aid for Research Activity Start-up 23840044,
Grant-in-Aid for Specially Promoted Research 22000005,
Grant-in-Aid for Young Scientists (A) 25707012, 
Grant-in-Aid Scientific Research (C) 21540240, 
the Global COE Program ``The Next Generation of Physics, Spun from
Universality and Emergence'', and Grants for Excellent Graduate Schools, from the Ministry of Education,
Culture, Sports, Science and Technology (MEXT) of Japan.
%We especially thank the anonymous referee for constructive comments which have significantly improved the manuscript.

\newpage

%%%%%%%%%%%%%%%%%%%%bibliography%%%%%%%%%%%%%%%%%

\newpage

%%%%%%%%%%%%%%%%%%%%figure%%%%%%%%%%%%%%%%%%%%%%

%\begin{landscape}
\begin{table}[tb]
\footnotesize
\caption{Distinctive polarized sources.}
\label{tab:catalogue_I}
\begin{center}
\begin{tabular}{cccccc} \hline
ID & $l$ [deg] & $b$ [deg] & m$_{K_S}$ & ($Q/I$, $U/I$) & $p_{{\rm int}}$[\%], $\theta_{{\rm int}}$ [deg] $^a$ \\ \hline
1 & -0.36587 & 0.00716 & 13.05 & (-0.0057$\pm$0.0052, -0.0232$\pm$0.0033) & 	(2.36$\pm$0.35, 128.2$\pm$9.3) \\  
2 & -0.32143 & -0.00002 & 13.81 & (0.0397$\pm$0.0063, 0.0978$\pm$0.0073) & 	(10.53$\pm$0.72, 34.0$\pm$14.1) \\  
3 & -0.31493 & 0.00294 & 13.42 & (0.1088$\pm$0.0068, 0.0549$\pm$0.0051) & 	(12.17$\pm$0.65, 13.4$\pm$16.3) \\  
4 & -0.31323 & 0.03271 & 14.23 & (0.1116$\pm$0.0076, 0.0540$\pm$0.0063) & 	(12.37$\pm$0.74, 12.9$\pm$15.9) \\  
5 & -0.26136 & -0.02096 & 13.83 & (-0.0057$\pm$0.0119, 0.0930$\pm$0.0100) & 	(9.26$\pm$1.00, 46.8$\pm$2.5) \\  
6 & -0.50482 & -0.01071 & 13.48 & (-0.0534$\pm$0.0049, -0.0247$\pm$0.0076) & 	(5.86$\pm$0.55, 102.4$\pm$15.4) \\  
7 & -0.54418 & -0.04250 & 9.76 & (0.0046$\pm$0.0038, -0.1065$\pm$0.0062) & 	(10.65$\pm$0.61, 136.2$\pm$1.7) \\  
8 & -0.54252 & -0.04275 & 13.46 & (-0.0403$\pm$0.0090, -0.0524$\pm$0.0062) & 	(6.57$\pm$0.74, 116.2$\pm$19.6) \\  
9 & -0.53515 & 0.02455 & 11.92 & (0.0918$\pm$0.0036, -0.0499$\pm$0.0061) & 	(10.44$\pm$0.43, 165.7$\pm$17.0) \\  
10 & -0.70771 & -0.03151 & 14.48 & (-0.0626$\pm$0.0113, -0.0561$\pm$0.0109) & 	(8.33$\pm$1.11, 110.9$\pm$20.1) \\  
11 & -0.69961 & -0.00496 & 12.47 & (-0.0067$\pm$0.0017, -0.0489$\pm$0.0027) & 	(4.93$\pm$0.27, 131.1$\pm$5.5) \\  
12 & -0.69310 & 0.01078 & 14.22 & (0.0721$\pm$0.0075, -0.0918$\pm$0.0041) & 	(11.66$\pm$0.57, 154.1$\pm$19.7) \\  
13 & -0.64849 & -0.00737 & 13.19 & (0.0394$\pm$0.0038, -0.0774$\pm$0.0044) & 	(8.67$\pm$0.43, 148.5$\pm$16.4) \\  
14 & -0.69671 & 0.04847 & 12.16 & (0.0616$\pm$0.0031, -0.0608$\pm$0.0027) & 	(8.65$\pm$0.29, 157.7$\pm$20.3) \\  
15 & -0.70104 & 0.08365 & 13.57 & (0.1199$\pm$0.0075, 0.0328$\pm$0.0074) & 	(12.41$\pm$0.75, 7.6$\pm$10.3) \\  
16 & -0.72759 & 0.01026 & 13.61 & (0.0807$\pm$0.0085, 0.0752$\pm$0.0053) & 	(11.01$\pm$0.72, 21.5$\pm$20.2) \\  
17 & -0.72551 & 0.01208 & 11.34 & (0.0764$\pm$0.0010, 0.0597$\pm$0.0012) & 	(9.69$\pm$0.10, 19.0$\pm$19.7) \\  
18 & -0.71920 & 0.00989 & 12.45 & (0.0040$\pm$0.0022, 0.0709$\pm$0.0018) & 	(7.10$\pm$0.18, 43.4$\pm$2.3) \\  
19 & -0.84658 & -0.01125 & 11.09 & (0.0132$\pm$0.0014, -0.0569$\pm$0.0018) & 	(5.84$\pm$0.18, 141.5$\pm$8.9) \\  
20 & -1.03901 & 0.09324 & 14.02 & (0.0975$\pm$0.0100, 0.0348$\pm$0.0119) & 	(10.30$\pm$1.02, 9.8$\pm$12.8) \\  
21 & -1.01488 & 0.08482 & 13.19 & (0.0938$\pm$0.0050, 0.0261$\pm$0.0072) & 	(9.73$\pm$0.52, 7.8$\pm$10.5) \\  
22 & -0.99319 & 0.11401 & 9.50 & (0.0083$\pm$0.0020, -0.0542$\pm$0.0027) & 	(5.48$\pm$0.27, 139.4$\pm$6.1) \\  
23 & -1.01675 & 0.12500 & 13.05 & (0.0056$\pm$0.0059, -0.0555$\pm$0.0075) & 	(5.52$\pm$0.75, 137.9$\pm$4.0) \\  
24 & -1.08127 & -0.03959 & 13.71 & (-0.0605$\pm$0.0078, 0.0539$\pm$0.0076) & 	(8.07$\pm$0.77, 69.1$\pm$20.1) \\  
25 & -1.08312 & -0.03312 & 13.78 & (0.0013$\pm$0.0059, 0.1211$\pm$0.0059) & 	(12.09$\pm$0.59, 44.7$\pm$0.4) \\  
26 & -1.07916 & -0.02981 & 14.36 & (0.0945$\pm$0.0093, 0.0062$\pm$0.0082) & 	(9.43$\pm$0.93, 1.9$\pm$2.6) \\  
27 & -1.07947 & -0.02715 & 13.44 & (0.0792$\pm$0.0055, -0.0170$\pm$0.0053) & 	(8.08$\pm$0.55, 173.9$\pm$8.3) \\  
28 & -1.12592 & 0.05842 & 13.42 & (0.0568$\pm$0.0059, 0.0734$\pm$0.0051) & 	(9.27$\pm$0.54, 26.1$\pm$19.6) \\  
29 & -1.06921 & -0.02958 & 12.05 & (0.0551$\pm$0.0023, -0.1178$\pm$0.0031) & 	(13.00$\pm$0.30, 147.5$\pm$15.5) \\  
30 & -1.09349 & 0.06670 & 13.09 & (0.0990$\pm$0.0040, 0.0294$\pm$0.0047) & 	(10.32$\pm$0.41, 8.3$\pm$11.1) \\  
31 & -1.08597 & 0.06459 & 13.59 & (0.0980$\pm$0.0044, 0.0237$\pm$0.0041) & 	(10.07$\pm$0.44, 6.8$\pm$9.3) \\  
32 & -1.04325 & 0.10097 & 12.53 & (0.0688$\pm$0.0039, -0.0008$\pm$0.0061) & 	(6.87$\pm$0.39, 179.7$\pm$0.4) \\  
33 & -1.05950 & 0.14882 & 14.26 & (-0.1001$\pm$0.0189, -0.0209$\pm$0.0193) & 	(10.05$\pm$1.89, 95.9$\pm$8.1) \\  
34 & -1.32709 & -0.04970 & 14.48 & (0.0550$\pm$0.0170, 0.1313$\pm$0.0201) & 	(14.10$\pm$1.97, 33.6$\pm$14.4) \\  
35 & -1.35976 & 0.01652 & 13.23 & (0.0596$\pm$0.0053, 0.0568$\pm$0.0031) & 	(8.22$\pm$0.44, 21.8$\pm$20.2) \\  

\end{tabular}
\end{center}
$^a$ To calculate intrinsic polarization, 
we estimate typical interstellar polarization from 1000 stars around distinctive polarized source in question, 
and subtract the interstellar polarization from observed polarization of distinctive polarized source in question. \\
\end{table}

\newpage
\setcounter{table}{0}
\begin{table}[tb]
\footnotesize
\caption{(Continued.)}
\label{tab:catalogue_II}
\begin{center}
\begin{tabular}{cccccc} \hline
ID & $l$ [deg] & $b$ [deg] & m$_{K_S}$ & ($Q/I$, $U/I$) & $p_{{\rm int}}$[\%], $\theta_{{\rm int}}$ [deg] \\ \hline
36 & -1.29967 & 0.04076 & 14.33 & (0.0804$\pm$0.0045, 0.0064$\pm$0.0098) & 	(8.06$\pm$0.45, 2.3$\pm$3.2) \\  
37 & -1.44430 & 0.08480 & 12.28 & (-0.0056$\pm$0.0037, 0.0799$\pm$0.0047) & 	(8.00$\pm$0.47, 47.0$\pm$2.8) \\  
38 & 1.00239 & -0.01601 & 13.20 & (-0.0032$\pm$0.0042, 0.0959$\pm$0.0058) & 	(9.58$\pm$0.58, 45.9$\pm$1.3) \\  
39 & 1.01134 & -0.02427 & 13.83 & (0.0196$\pm$0.0061, 0.1066$\pm$0.0069) & 	(10.82$\pm$0.69, 39.8$\pm$7.2) \\  
40 & 0.99038 & 0.01079 & 13.81 & (0.0728$\pm$0.0054, 0.0822$\pm$0.0070) & 	(10.96$\pm$0.64, 24.2$\pm$20.1) \\  
41 & 1.01060 & -0.00624 & 14.40 & (-0.0818$\pm$0.0095, 0.1706$\pm$0.0145) & 	(18.87$\pm$1.37, 57.8$\pm$15.8) \\  
42 & 0.94197 & 0.01020 & 14.38 & (0.0460$\pm$0.0166, 0.1500$\pm$0.0129) & 	(15.63$\pm$1.32, 36.5$\pm$11.4) \\  
43 & 1.11969 & -0.01968 & 13.41 & (-0.0412$\pm$0.0039, 0.0926$\pm$0.0050) & 	(10.13$\pm$0.49, 57.0$\pm$15.1) \\  
44 & 1.12922 & -0.01782 & 14.04 & (-0.0453$\pm$0.0075, 0.1172$\pm$0.0100) & 	(12.53$\pm$0.97, 55.6$\pm$13.6) \\  
45 & 1.13035 & -0.00307 & 14.49 & (-0.0603$\pm$0.0087, 0.0840$\pm$0.0110) & 	(10.29$\pm$1.03, 62.8$\pm$19.2) \\  
46 & 1.13639 & -0.00017 & 14.46 & (-0.0617$\pm$0.0091, 0.0863$\pm$0.0095) & 	(10.57$\pm$0.94, 62.8$\pm$19.2) \\  
47 & 1.19906 & -0.05575 & 11.56 & (-0.0410$\pm$0.0018, -0.0217$\pm$0.0022) & 	(4.64$\pm$0.19, 103.9$\pm$16.7) \\  
48 & 1.19685 & -0.03102 & 11.59 & (0.0436$\pm$0.0013, -0.0437$\pm$0.0027) & 	(6.17$\pm$0.21, 157.5$\pm$20.3) \\  
49 & 1.23144 & -0.05336 & 12.55 & (-0.0756$\pm$0.0021, 0.0371$\pm$0.0018) & 	(8.42$\pm$0.21, 76.9$\pm$16.0) \\  
50 & 1.30546 & 0.04850 & 13.54 & (-0.0339$\pm$0.0067, 0.0870$\pm$0.0041) & 	(9.32$\pm$0.45, 55.7$\pm$13.7) \\  
51 & 1.42381 & -0.05024 & 14.43 & (-0.0012$\pm$0.0077, 0.0834$\pm$0.0096) & 	(8.28$\pm$0.96, 45.4$\pm$0.6) \\  
52 & 1.47688 & 0.00674 & 13.17 & (-0.0303$\pm$0.0028, 0.0649$\pm$0.0033) & 	(7.16$\pm$0.32, 57.5$\pm$15.5) \\  
53 & 1.44361 & 0.06872 & 12.22 & (-0.0014$\pm$0.0021, 0.0638$\pm$0.0006) & 	(6.38$\pm$0.06, 45.6$\pm$0.9) \\  
54 & 1.46587 & 0.05014 & 13.89 & (-0.0759$\pm$0.0061, 0.0328$\pm$0.0048) & 	(8.25$\pm$0.59, 78.3$\pm$14.8) \\  
55 & 1.46881 & 0.04728 & 13.36 & (-0.0689$\pm$0.0041, 0.0396$\pm$0.0053) & 	(7.94$\pm$0.44, 75.1$\pm$17.5) \\  
56 & 1.46959 & 0.04754 & 13.37 & (-0.0444$\pm$0.0028, 0.0410$\pm$0.0054) & 	(6.03$\pm$0.42, 68.6$\pm$20.2) \\  
57 & 1.48281 & 0.03689 & 12.12 & (-0.0469$\pm$0.0016, -0.0151$\pm$0.0018) & 	(4.93$\pm$0.16, 98.9$\pm$11.8) \\  
58 & 1.52982 & -0.06968 & 11.64 & (0.0213$\pm$0.0033, 0.0770$\pm$0.0018) & 	(7.98$\pm$0.19, 37.3$\pm$10.4) \\  
59 & 1.52653 & -0.03607 & 13.31 & (-0.0468$\pm$0.0034, 0.0177$\pm$0.0031) & 	(5.00$\pm$0.33, 79.6$\pm$13.4) \\  
60 & 1.54785 & -0.04246 & 12.41 & (-0.0538$\pm$0.0022, -0.0017$\pm$0.0014) & 	(5.38$\pm$0.22, 90.9$\pm$1.3) \\  
61 & 1.50826 & 0.03478 & 11.27 & (-0.0270$\pm$0.0018, -0.0209$\pm$0.0014) & 	(3.41$\pm$0.17, 108.9$\pm$19.6) \\  
62 & 1.54194 & 0.00720 & 13.30 & (-0.0159$\pm$0.0053, -0.0573$\pm$0.0028) & 	(5.94$\pm$0.30, 127.2$\pm$10.5) \\  
63 & 1.54581 & 0.00758 & 13.02 & (-0.0691$\pm$0.0026, 0.0495$\pm$0.0037) & 	(8.49$\pm$0.30, 72.2$\pm$19.2) \\  
64 & 1.57701 & -0.03735 & 13.74 & (-0.0806$\pm$0.0075, -0.0483$\pm$0.0075) & 	(9.36$\pm$0.75, 105.5$\pm$17.9) \\  
65 & 1.57533 & -0.01951 & 12.14 & (0.0301$\pm$0.0033, 0.0594$\pm$0.0019) & 	(6.65$\pm$0.22, 31.6$\pm$16.3) \\  
66 & 1.56381 & 0.00990 & 13.53 & (-0.0120$\pm$0.0036, 0.0696$\pm$0.0045) & 	(7.05$\pm$0.45, 49.9$\pm$6.8) \\  
67 & 1.55684 & 0.02260 & 13.01 & (0.0383$\pm$0.0034, 0.0557$\pm$0.0036) & 	(6.75$\pm$0.35, 27.7$\pm$18.9) \\  
68 & 1.58666 & -0.01681 & 12.94 & (0.0298$\pm$0.0039, 0.0593$\pm$0.0018) & 	(6.63$\pm$0.24, 31.6$\pm$16.3) \\  
69 & 1.56265 & 0.02780 & 13.85 & (0.0693$\pm$0.0061, 0.0385$\pm$0.0048) & 	(7.91$\pm$0.58, 14.5$\pm$17.2) \\  
70 & 1.58672 & -0.00505 & 12.92 & (0.0708$\pm$0.0025, 0.0532$\pm$0.0029) & 	(8.86$\pm$0.26, 18.5$\pm$19.5) \\  
71 & 1.60044 & -0.01318 & 13.05 & (0.0412$\pm$0.0034, 0.0538$\pm$0.0044) & 	(6.77$\pm$0.41, 26.3$\pm$19.6) \\  
72 & 1.60231 & -0.00553 & 14.48 & (-0.0542$\pm$0.0073, -0.0043$\pm$0.0072) & 	(5.39$\pm$0.73, 92.2$\pm$3.2) \\  
73 & 1.60753 & -0.00844 & 13.48 & (-0.0336$\pm$0.0048, 0.0614$\pm$0.0045) & 	(6.98$\pm$0.46, 59.3$\pm$17.1) \\  
74 & 1.63677 & -0.06471 & 14.12 & (-0.0122$\pm$0.0199, 0.0874$\pm$0.0091) & 	(8.77$\pm$0.95, 49.0$\pm$5.6) \\  
75 & 1.66162 & -0.03622 & 13.37 & (0.0352$\pm$0.0039, -0.0433$\pm$0.0049) & 	(5.56$\pm$0.46, 154.5$\pm$19.8) \\   

\end{tabular}
\end{center}
\end{table}

\newpage
\setcounter{table}{0}
\begin{table}[tb]
\footnotesize
\caption{(Continued.)}
\label{tab:catalogue_III}
\begin{center}
\begin{tabular}{cccccc} \hline
ID & $l$ [deg] & $b$ [deg] & m$_{K_S}$ & ($Q/I$, $U/I$) & $p_{{\rm int}}$[\%], $\theta_{{\rm int}}$ [deg] \\ \hline
76 & 1.66031 & -0.00979 & 13.00 & (0.0833$\pm$0.0035, -0.0160$\pm$0.0026) & 	(8.48$\pm$0.35, 174.6$\pm$7.5) \\  
77 & 1.68901 & -0.04740 & 13.37 & (0.0551$\pm$0.0070, 0.0408$\pm$0.0035) & 	(6.83$\pm$0.60, 18.3$\pm$19.4) \\  
78 & 1.67463 & 0.03332 & 12.35 & (0.0423$\pm$0.0026, 0.0445$\pm$0.0023) & 	(6.14$\pm$0.24, 23.2$\pm$20.2) \\  
79 & 1.66785 & 0.06775 & 12.90 & (-0.0249$\pm$0.0027, 0.0433$\pm$0.0025) & 	(4.99$\pm$0.25, 59.9$\pm$17.5) \\  
80 & 1.63009 & -0.05215 & 13.39 & (0.0130$\pm$0.0058, 0.0867$\pm$0.0053) & 	(8.76$\pm$0.53, 40.7$\pm$6.0) \\ 
81 & 1.63103 & -0.03721 & 12.97 & (0.0067$\pm$0.0034, 0.0904$\pm$0.0031) & 	(9.06$\pm$0.31, 42.9$\pm$3.0) \\  
82 & 1.60188 & 0.02516 & 13.63 & (0.0779$\pm$0.0061, 0.0249$\pm$0.0066) & 	(8.15$\pm$0.62, 8.9$\pm$11.8) \\  
83 & 1.65283 & 0.01719 & 12.39 & (0.0899$\pm$0.0193, 0.1798$\pm$0.0418) & 	(19.73$\pm$3.83, 31.7$\pm$16.2) \\  
84 & 1.77185 & -0.00050 & 14.36 & (0.0705$\pm$0.0088, -0.0398$\pm$0.0098) & 	(8.05$\pm$0.91, 165.3$\pm$17.3) \\  
85 & 1.77551 & 0.00168 & 13.09 & (0.0749$\pm$0.0045, 0.0019$\pm$0.0066) & 	(7.48$\pm$0.45, 0.7$\pm$1.0) \\  
86 & 1.92631 & 0.03101 & 12.62 & (0.0310$\pm$0.0030, 0.0992$\pm$0.0027) & 	(10.39$\pm$0.28, 36.3$\pm$11.5) \\  
87 & 2.12453 & -0.00816 & 13.69 & (0.0924$\pm$0.0067, 0.0418$\pm$0.0065) & 	(10.12$\pm$0.67, 12.2$\pm$15.2) \\  
88 & 2.13293 & 0.02697 & 13.47 & (0.0248$\pm$0.0046, 0.0822$\pm$0.0045) & 	(8.58$\pm$0.45, 36.6$\pm$11.2) \\  
89 & 2.14388 & 0.02429 & 13.86 & (0.0573$\pm$0.0060, 0.0823$\pm$0.0064) & 	(10.01$\pm$0.62, 27.6$\pm$19.0) \\  
90 & 2.14307 & 0.04332 & 13.56 & (0.0019$\pm$0.0064, -0.0408$\pm$0.0045) & 	(4.06$\pm$0.45, 136.3$\pm$1.8) \\  
91 & 2.16228 & 0.01616 & 12.99 & (0.0670$\pm$0.0031, 0.0526$\pm$0.0030) & 	(8.52$\pm$0.31, 19.1$\pm$19.7) \\  
92 & 2.20481 & 0.02567 & 13.85 & (0.1080$\pm$0.0064, 0.0353$\pm$0.0061) & 	(11.35$\pm$0.63, 9.1$\pm$12.0) \\  
93 & 2.21070 & 0.03195 & 13.29 & (0.0855$\pm$0.0047, 0.0377$\pm$0.0054) & 	(9.33$\pm$0.48, 11.9$\pm$15.0) \\  
94 & 2.22717 & 0.03583 & 13.35 & (0.0494$\pm$0.0044, 0.0858$\pm$0.0047) & 	(9.89$\pm$0.46, 30.0$\pm$17.5) \\  
95 & 2.24174 & 0.03996 & 14.25 & (0.0028$\pm$0.0065, 0.1116$\pm$0.0112) & 	(11.10$\pm$1.12, 44.3$\pm$1.0) \\  
96 & 2.19070 & 0.02931 & 14.07 & (0.0621$\pm$0.0070, -0.0496$\pm$0.0105) & 	(7.90$\pm$0.85, 160.7$\pm$19.8) \\  
97 & 2.32129 & 0.02894 & 13.37 & (0.0340$\pm$0.0058, 0.0967$\pm$0.0039) & 	(10.24$\pm$0.41, 35.3$\pm$12.7) \\  
98 & 2.31065 & 0.05186 & 11.51 & (-0.0615$\pm$0.0026, 0.0069$\pm$0.0036) & 	(6.18$\pm$0.26, 86.8$\pm$4.5) \\  
99 & 2.38885 & 0.00187 & 11.17 & (0.0409$\pm$0.0028, 0.0509$\pm$0.0017) & 	(6.52$\pm$0.22, 25.6$\pm$19.8) \\  
100 & 2.43841 & -0.03000 & 14.05 & (-0.0652$\pm$0.0068, 0.0209$\pm$0.0068) & 	(6.82$\pm$0.68, 81.1$\pm$11.8) \\  
101 & 2.77446 & -0.02202 & 14.06 & (0.0067$\pm$0.0065, -0.0662$\pm$0.0049) & 	(6.63$\pm$0.49, 137.9$\pm$4.1) \\  
102 & 2.77709 & -0.01002 & 13.63 & (0.0521$\pm$0.0030, -0.0545$\pm$0.0066) & 	(7.52$\pm$0.52, 156.8$\pm$20.2) \\  
103 & 2.78503 & -0.01580 & 10.60 & (0.0256$\pm$0.0008, -0.0586$\pm$0.0011) & 	(6.39$\pm$0.11, 146.8$\pm$14.9) \\  
104 & 2.78360 & -0.01150 & 12.17 & (0.0490$\pm$0.0016, -0.0478$\pm$0.0018) & 	(6.84$\pm$0.17, 157.9$\pm$20.3) \\  
105 & 2.78312 & -0.00583 & 13.53 & (0.0502$\pm$0.0042, -0.0513$\pm$0.0050) & 	(7.16$\pm$0.46, 157.2$\pm$20.3) \\  
106 & 2.78690 & -0.01112 & 12.21 & (0.0466$\pm$0.0021, -0.0622$\pm$0.0019) & 	(7.77$\pm$0.20, 153.4$\pm$19.4) \\  
107 & 2.81163 & -0.01477 & 13.54 & (0.0804$\pm$0.0049, -0.0501$\pm$0.0042) & 	(9.46$\pm$0.47, 164.0$\pm$18.2) \\  
108 & 2.85714 & -0.00776 & 12.16 & (0.0873$\pm$0.0022, -0.0282$\pm$0.0015) & 	(9.17$\pm$0.21, 171.0$\pm$11.9) \\  
109 & 2.85853 & -0.00807 & 14.29 & (0.0770$\pm$0.0071, -0.0423$\pm$0.0069) & 	(8.76$\pm$0.70, 165.6$\pm$17.1) \\  
110 & 2.86513 & -0.00967 & 13.14 & (0.0871$\pm$0.0035, -0.0048$\pm$0.0032) & 	(8.72$\pm$0.35, 178.4$\pm$2.2) \\  
111 & 2.89945 & -0.03906 & 13.68 & (0.0552$\pm$0.0037, 0.0626$\pm$0.0036) & 	(8.33$\pm$0.36, 24.3$\pm$20.1) \\  
112 & 2.87477 & 0.00756 & 12.55 & (-0.0022$\pm$0.0020, -0.0490$\pm$0.0017) & 	(4.90$\pm$0.17, 133.7$\pm$1.9) \\    
\hline

\end{tabular}
\end{center}
\end{table}
%\end{landscape}

\begin{table}[htbp]
\begin{center}
\caption{Matching results between the distinctive polarization sources and Spitzer data.}
\label{tab:matching}
\begin{tabular}{ccccc} \hline
[3.6]     & [4.5]     & [5.8]     & [8.0]     & [24]     \\ \hline
70 (63\%) $^a$ & 70 (63\%) & 68 (61\%) & 51 (46\%) & 11 (10\%) \\ \hline
\end{tabular}
\end{center}
$^a$ The number of the distinctive polarized sources matched with Spitzer data, 
and the fraction of the matched sources out of the 112 distinctive polarized sources.
\end{table}

\newpage
\begin{landscape}
\begin{table}[htbp]
\scriptsize
\begin{center}
\caption{Properties of YSO candidates}
\label{tab:YSO_candidates}
\begin{tabular}{ccccccc}\hline
No  & $H-K_S$ & ([3.6]-[4.5], [5.8]-[8.0]) & YSO class$^a$ & SED$^b$ & polarization type$^c$ & $<\Delta(H-K_S)>$ $^d$ \\ \hline
10 &  --   & (0.42, 0.97) & II   & normal & a & --    \\
11 &  2.07 & (1.47, 0.60) & II   & YSO    & a & -0.45 \\
13 &  4.00 & (1.18, 0.43) & I    & YSO    & a & 1.50  \\ 
14 &  4.00 & (0.97, 0.16) & III  & YSO    & a & 1.57  \\
19 &  3.39 & (1.21, 0.41) & I    & YSO    & a & 1.38  \\
21 &  3.85 & (0.53, 0.58) & II   & normal & a & 1.95  \\
29 &  3.68 & (1.29, 0.37) & I    & YSO    & a & 2.00  \\
40 &  4.01 & (1.47, 0.51) & I    & YSO    & a & 1.80  \\
98 &  3.93 & (0.77, 0.57) & II   & YSO    & b & 2.34  \\
112 & 3.97 & (--  , --  ) & --   & YSO    & a & 2.49  \\
\hline

\end{tabular}
\end{center}
$^a$ YSO classification in a [3.6]-[4.5] vs. [5.8]-[8.0] color-color diagram. 
I, II, and III represent Class I, II, and III YSOs. \\
$^b$  ``normal'' means that an only photsphere component can explain the SED of the source with fitting method of \citet{rob07}.
In contrast, ``YSO'' means that disk or/and envelope component is necessary. \\ 
$^c$ ``(a)'' and ``(b)'' means the types of panel (a) and (b) in the Figure \ref{fig:pol_diff}, respectively. \\
$^d$ $<\Delta(H-K_S)>$ is the difference of observed $H-K_S$ colors between distinctive polarized sources and the surrounding sources.
This is represented by $<(H-K_S)_{{\rm pol}}-(H-K_S)_{{\rm sur}}>$, 
where $(H-K_S)_{{\rm pol}}$ is the $H-K_S$ color of a distinctive polarized source, 
and $(H-K_S)_{{\rm sur}}$ is the $H-K_S$ color of surrounding sources.  
The positive value means that the distinctive polarized source is redder than surrounding sources.
\end{table}
\end{landscape}

%\hspace*{-2.5zw}
{\bf Appendix}

We show features of the 10 distinctive polarized sources classified as YSOs 
with color-color diagram or/and SED fitting.
All of 10 figures (Figure 18--27) contain four panels:
Upper left panel is a $K_S$-band image.
A yellow arrow indicates the position of a distinctive polarized source classified as a YSO.
Three blue circles represent $10''$, $20''$, and $30''$ from the distinctive polarized source.  
Red and green circles indicate possible distinctive polarized source(s) and other distinctive polarized source(s).
Upper right panel represents ``distance''  
between a distinctive polarized source and the average of surrounding sources in a $Q/I-U/I$ plane. 
Lower left panel is a $Q/I-U/I$ plane for 1000 surrounding sources around a distinctive polarized source, 
and black circle is the position of the distinctive polarized source in the $Q/I-U/I$ plane.  
Green and red plots are apart from the peak of $Q/I$ and $U/I$ of the surrounding 1000 sources 
by $5\,\sigma$ and 3--5\,$\sigma$, respectively.
Lower right panel is a histogram of the difference of $H-K_S$ colors 
between a distinctive polarized source and its surrounding sources.
Although we use the 1000 surrounding sources, 
some sources are not detected in the $H$ band,
and we remove them from the histogram.

\begin{figure}
\centering
\includegraphics[width=13cm,clip]{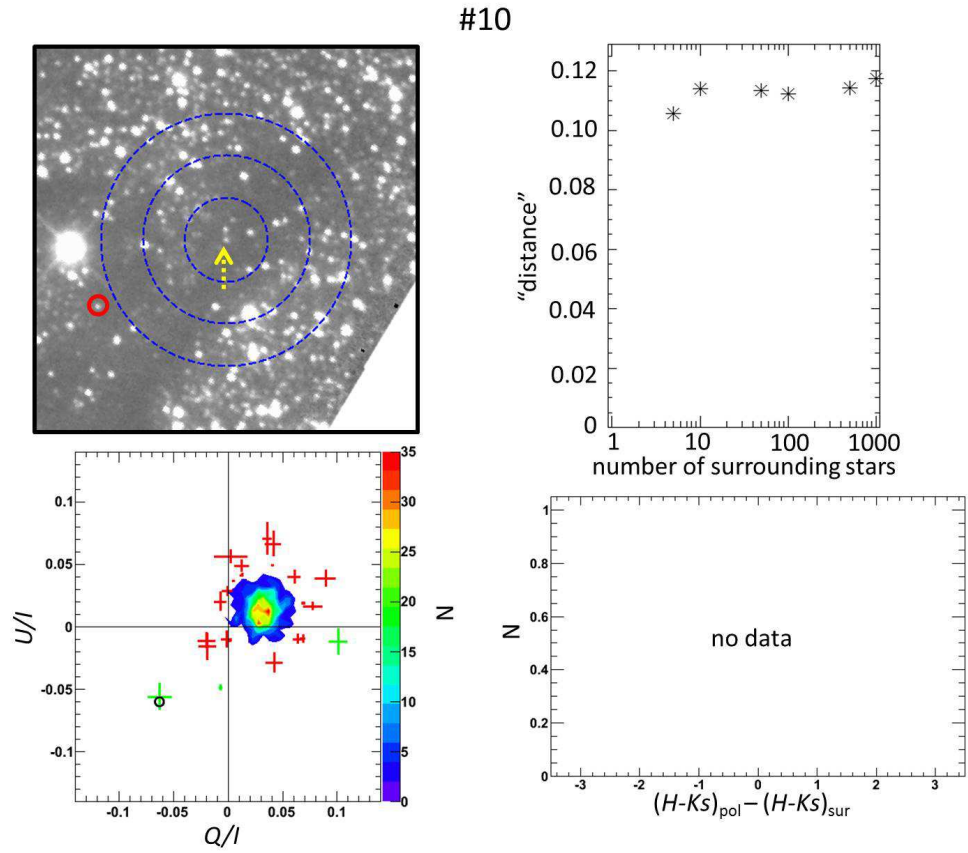}
\caption{The features of \#10.
\#10 does not have the $H$ band data.}
\end{figure}

\begin{figure}
\centering
\includegraphics[width=13cm,clip]{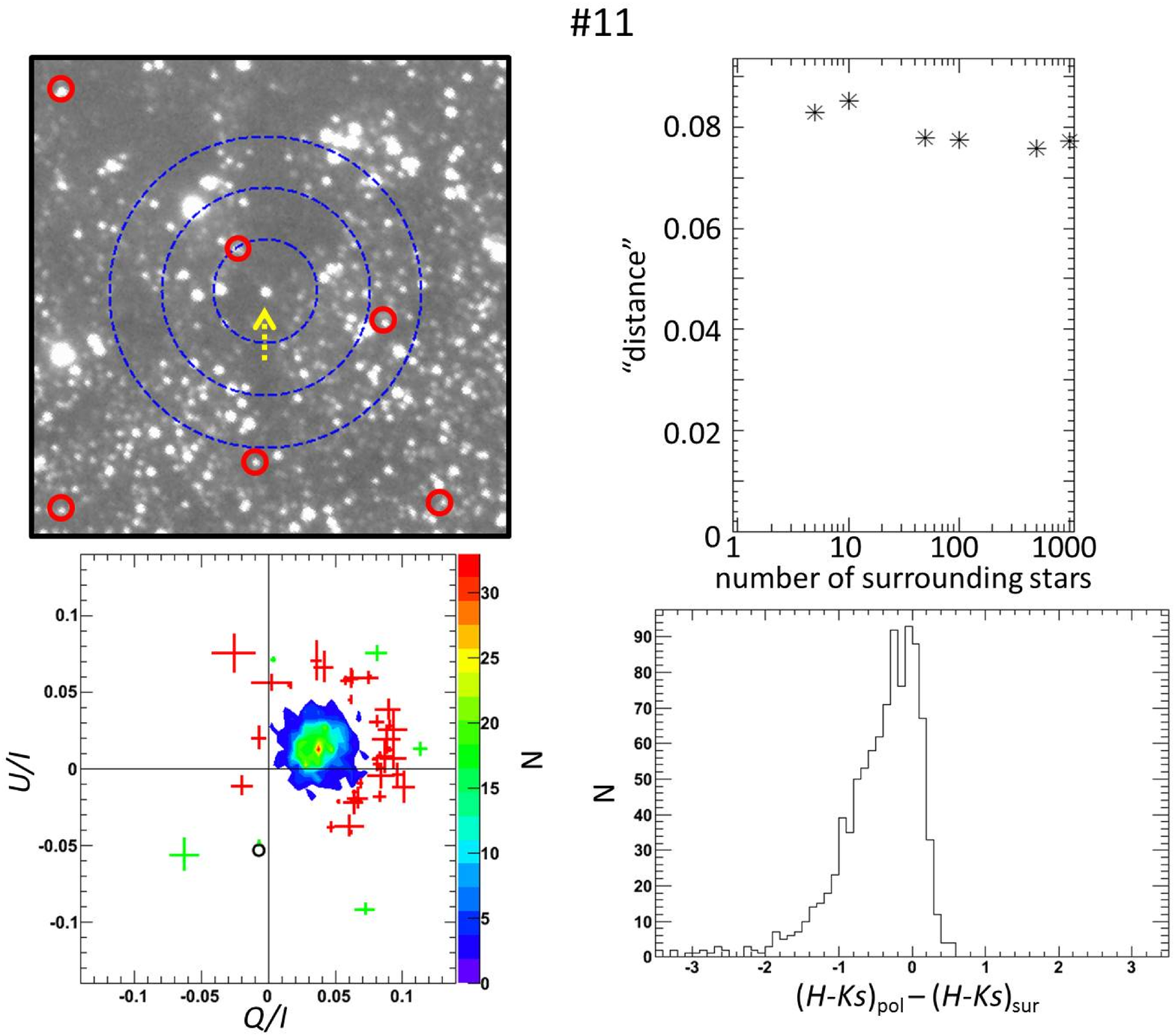}
\caption{The features of \#11.}
\end{figure}

\begin{figure}
\centering
\includegraphics[width=13cm,clip]{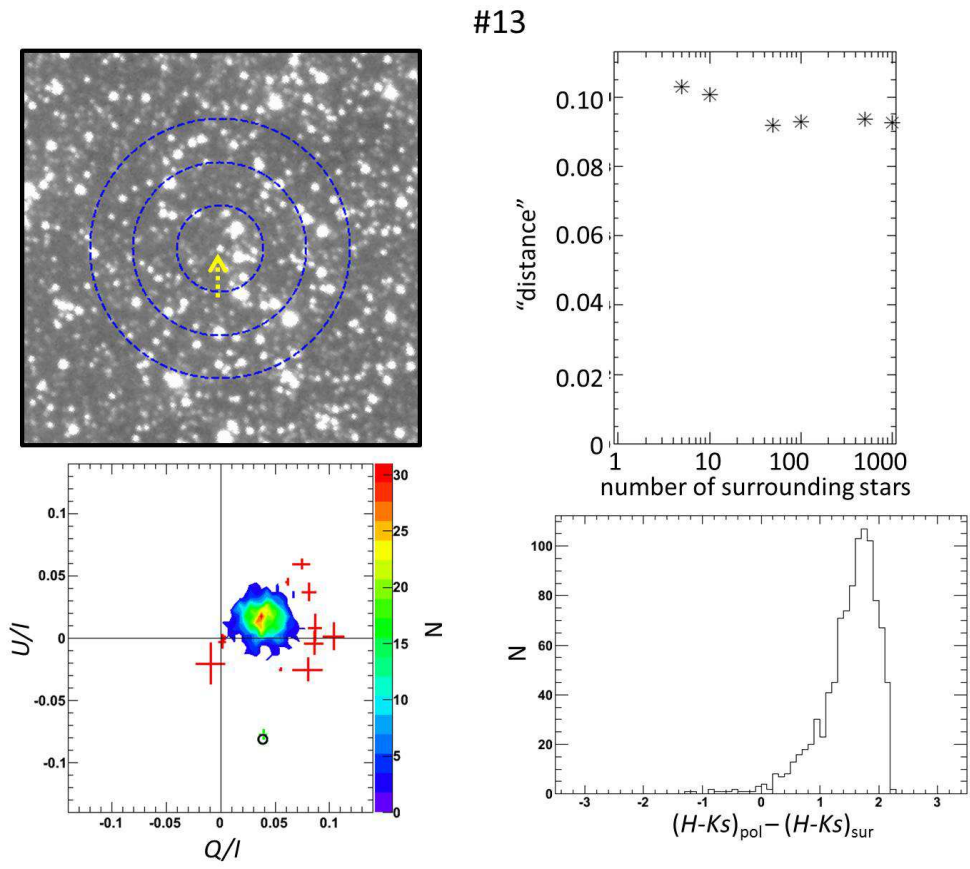}
\caption{The features of \#13.}
\end{figure} 

\begin{figure}
\centering
\includegraphics[width=13cm,clip]{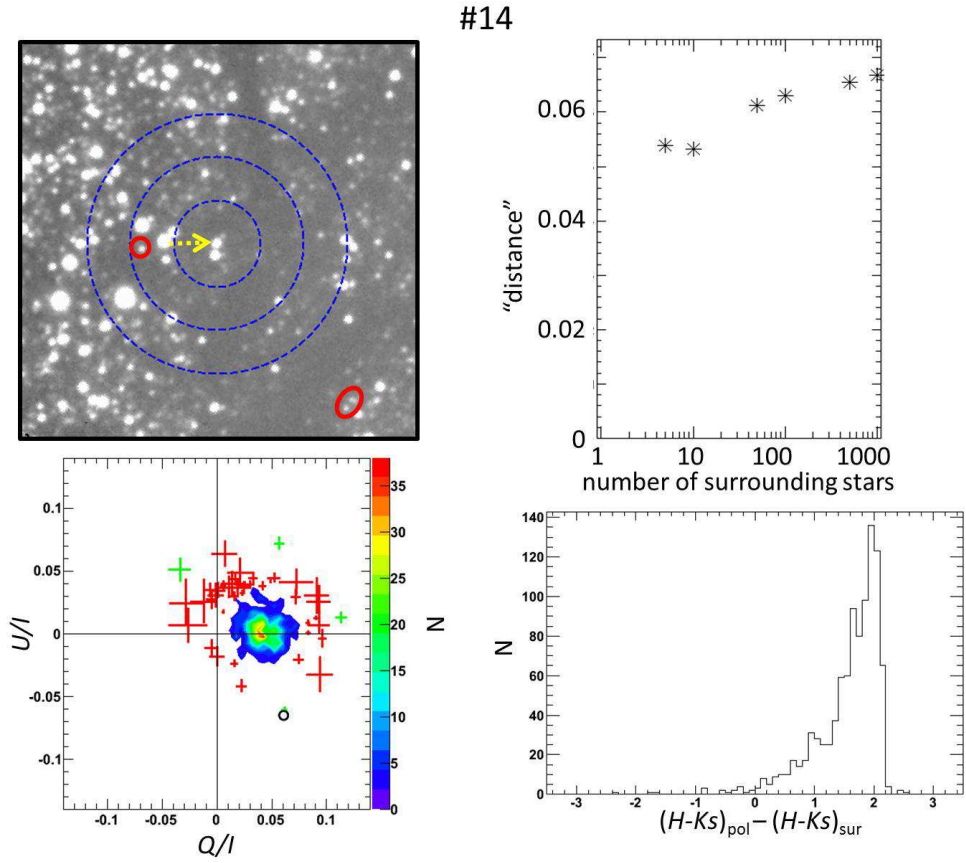}
\caption{The features of \#14.
The red ellipse in the lower left part of the $K_S$ band image circles two sources, 
and they are the possible distinctive polarized sources.
}
\end{figure}

\begin{figure}
\centering
\includegraphics[width=13cm,clip]{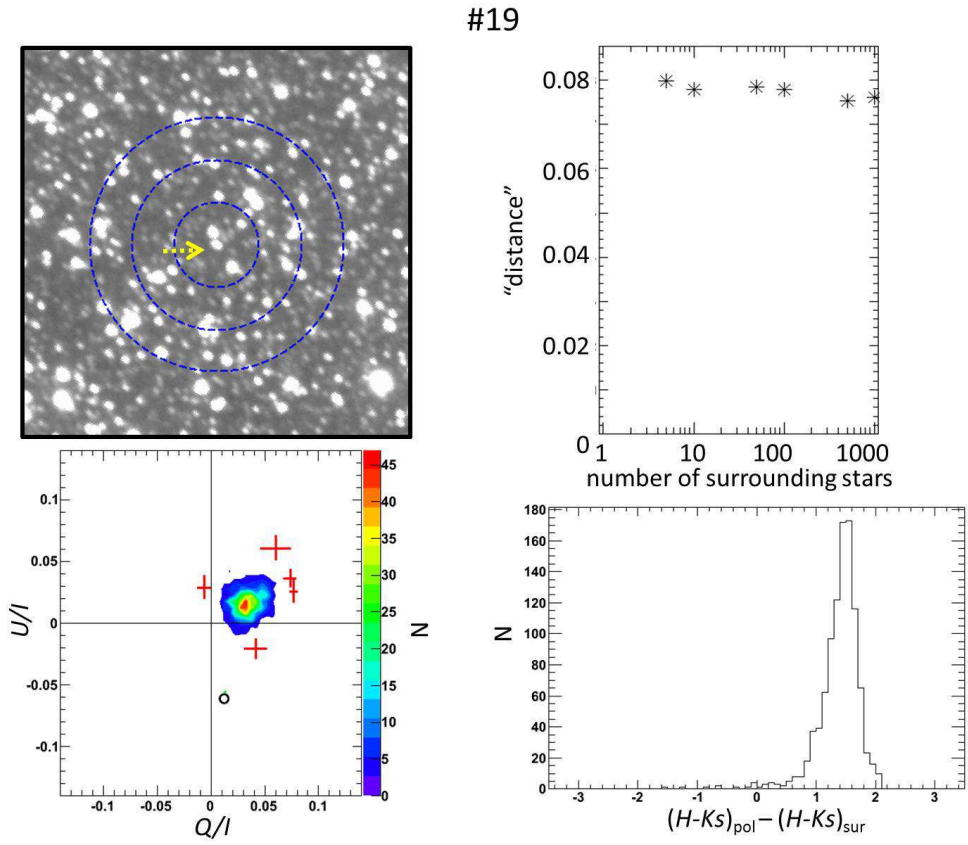}
\caption{The features of \#19.}
\end{figure}

\begin{figure}
\centering
\includegraphics[width=13cm,clip]{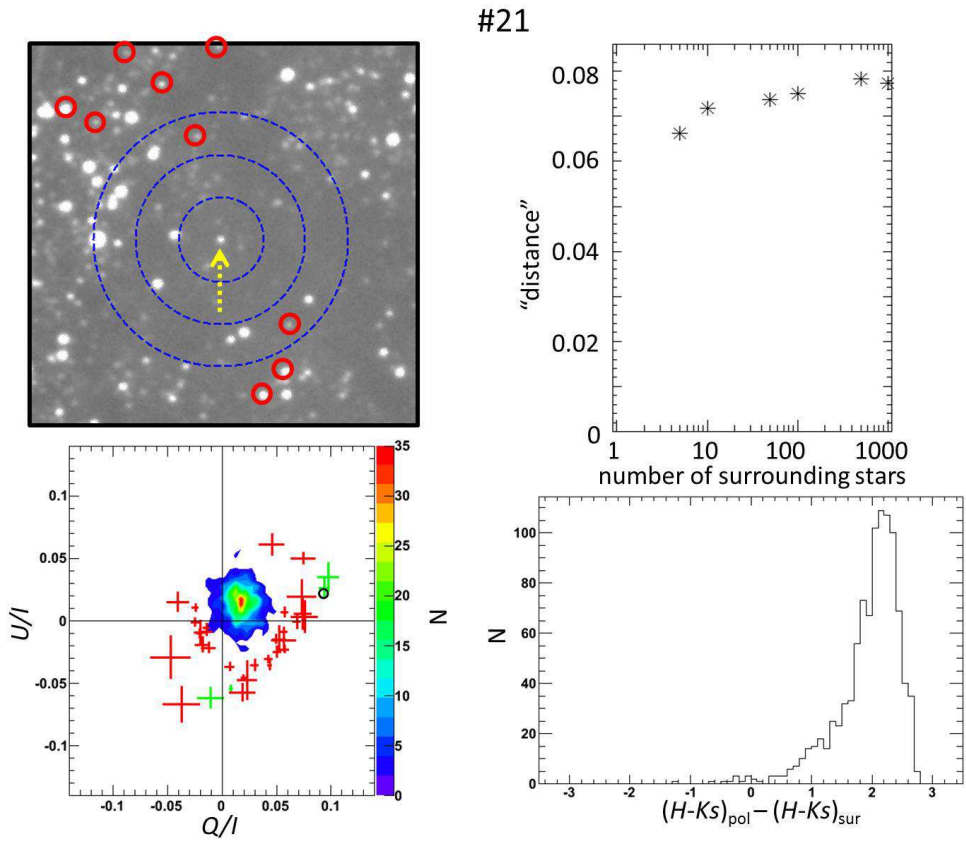}
\caption{The features of \#21.}
\end{figure}

\begin{figure}
\centering
\includegraphics[width=13cm,clip]{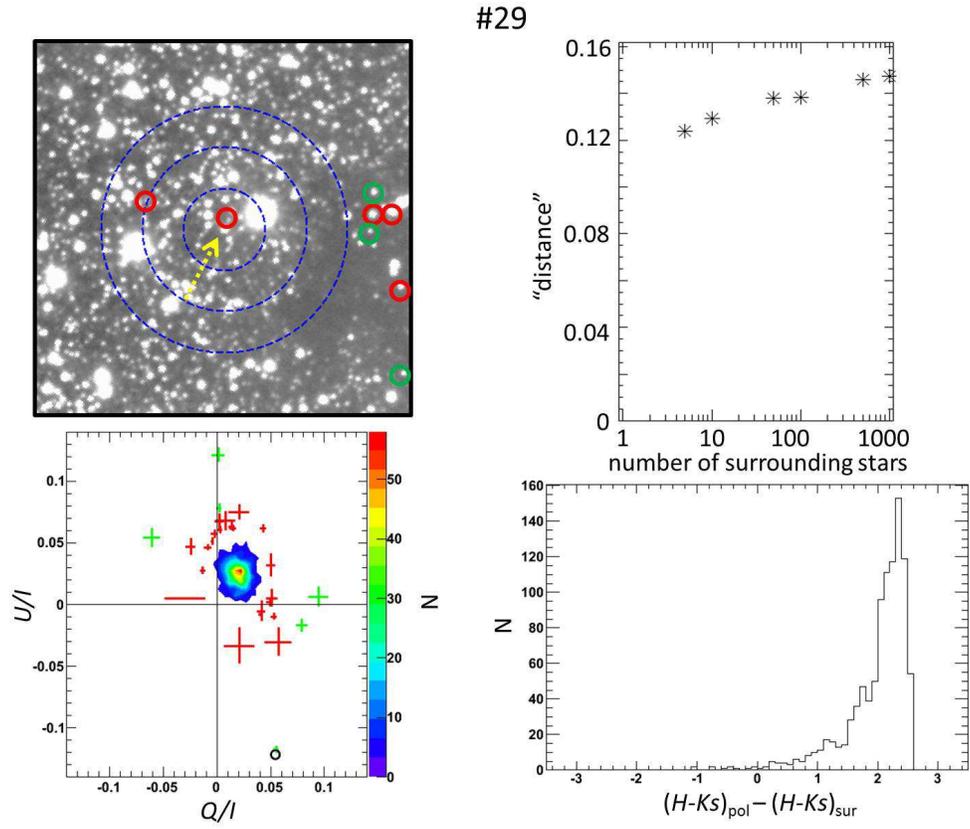}
\caption{The features of \#29.}
\end{figure}

\begin{figure}
\centering
\includegraphics[width=13cm,clip]{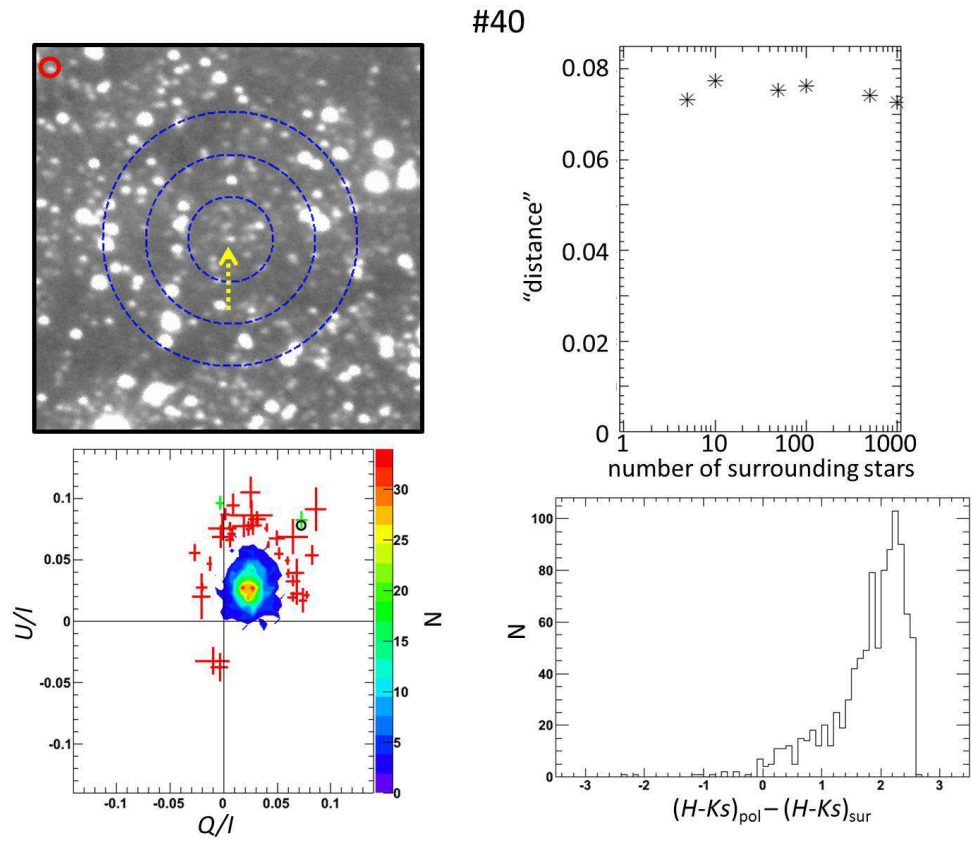}
\caption{The features of \#40.}
\end{figure}

\begin{figure}
\centering
\includegraphics[width=13cm,clip]{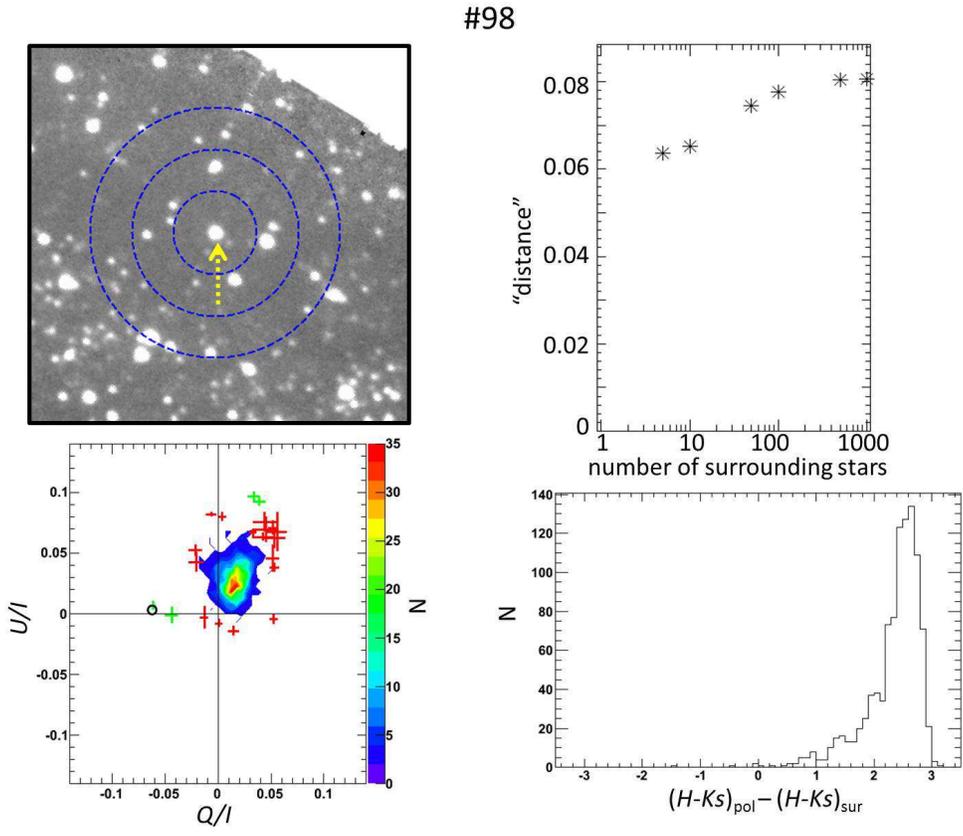}
\caption{The features of \#98.}
\end{figure}

\begin{figure}
\centering
\includegraphics[width=13cm,clip]{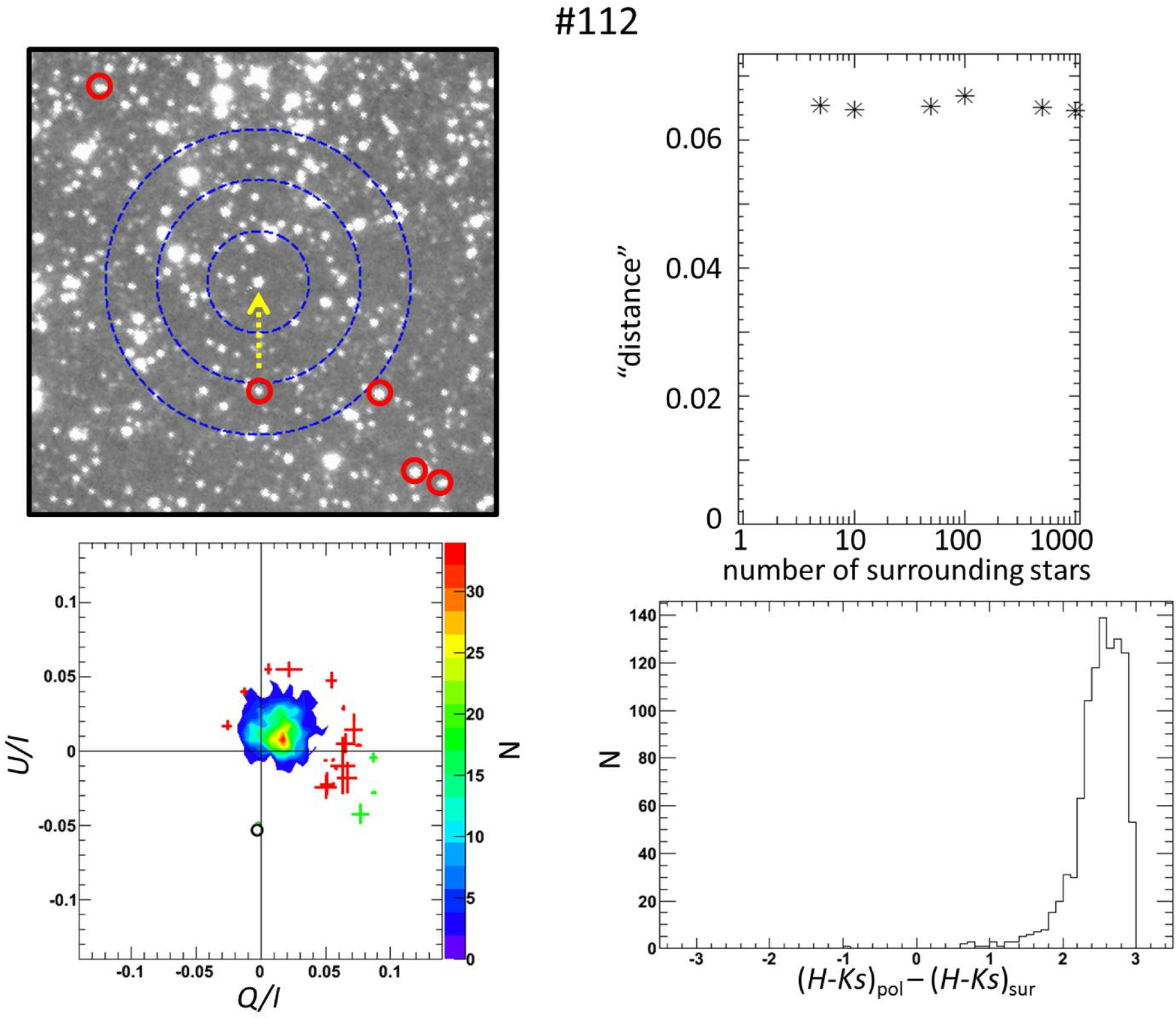}
\caption{The features of \#112.}
\end{figure}

\end{document}